\documentclass[nofootinbib, showpacs,superscriptaddress,10pt]{revtex4}
\usepackage{amssymb,amsmath}
\usepackage[dvips,final]{graphicx}
\usepackage{slashed}
\usepackage{url}
\usepackage{subfigure}
\usepackage{bbold}
\usepackage{textcomp}
\usepackage{verbatim}
\usepackage{dcolumn}
\usepackage[usenames]{color}
\usepackage{epsfig}
\input epsf.tex
\usepackage{bm}
\usepackage{amsthm}
\usepackage{braket}
\usepackage[title]{appendix}
\usepackage{amsopn}
\usepackage{ulem}
\usepackage{cancel}
\usepackage{hyperref}
\usepackage{soul}

\def\comment#1{}

\def\beq{\begin{equation}}
\def\eeq{\end{equation}}



\begin{document}

\title{CMB $V$ modes from photon-photon forward scattering revisited}

\author{Ahmad Hoseinpour}
\email[]{ahmad.hoseinpour@ph.iut.ac.ir}

\affiliation{Department of Physics, Isfahan University of Technology, Isfahan
84156-83111, Iran}
\affiliation{ICRANet-Isfahan, Isfahan University of Technology, Isfahan 84156-83111, Iran}

\author{Moslem Zarei}
\email[]{m.zarei@cc.iut.ac.ir}

\affiliation{Department of Physics, Isfahan University of Technology, Isfahan
84156-83111, Iran}
\affiliation{ICRANet-Isfahan, Isfahan University of Technology, Isfahan 84156-83111, Iran}
\affiliation{Dipartimento di Fisica e Astronomia \textquotedblleft G. Galilei\textquotedblright, Universit\`{a} degli Studi di Padova, via Marzolo
8, I-35131, Padova, Italy}
\affiliation{INFN Sezione di Padova, via Marzolo 8, I-35131, Padova, Italy}

\author{Giorgio Orlando}
\email[]{giorgio.orlando@phd.unipd.it}

\affiliation{Dipartimento di Fisica e Astronomia \textquotedblleft G. Galilei\textquotedblright, Universit\`{a} degli Studi di Padova, via Marzolo
8, I-35131, Padova, Italy}
\affiliation{INFN Sezione di Padova, via Marzolo 8, I-35131, Padova, Italy}

\author{Nicola Bartolo}
\email[]{nicola.bartolo@pd.infn.it}

\affiliation{Dipartimento di Fisica e Astronomia \textquotedblleft G. Galilei\textquotedblright, Universit\`{a} degli Studi di Padova, via Marzolo
8, I-35131, Padova, Italy}
\affiliation{INFN Sezione di Padova, via Marzolo 8, I-35131, Padova, Italy}
\affiliation{INAF-Osservatorio Astronomico di Padova, Vicolo dell'Osservatorio 5, I-35122 Padova, Italy}

\author{Sabino Matarrese}
\email[]{sabino.matarrese@pd.infn.it }

\affiliation{Dipartimento di Fisica e Astronomia \textquotedblleft G. Galilei\textquotedblright, Universit\`{a} degli Studi di Padova, via Marzolo
8, I-35131, Padova, Italy}
\affiliation{INFN Sezione di Padova, via Marzolo 8, I-35131, Padova, Italy}
\affiliation{INAF-Osservatorio Astronomico di Padova, Vicolo dell'Osservatorio 5, I-35122 Padova, Italy}
\affiliation{Gran Sasso Science Institute, viale F. Crispi 7, I-67100, L'Aquila, Italy}

\date{\today}

\begin{abstract}
Recent literature has shown that photon-photon forward scattering mediated by Euler-Heisenberg interactions may generate some amount of the circular polarization ($V$ modes) in the cosmic microwave background (CMB) photons. However, there is an apparent contradiction among the different references about the predicted level of the amplitude of this circular polarization. In this work, we will resolve this discrepancy by showing that with a quantum Boltzmann equation formalism, we obtain the same amount of circular polarization as using a geometrical approach that is based on the index of refraction of the cosmological medium. We will show that the expected amplitude of $V$ modes is expected to be $\approx 8$ orders of magnitude smaller than the amplitude of $E$-polarization modes that we actually observe in the CMB, thus confirming that it is going to be challenging to observe such a signature. Throughout the paper, we also develop a general method to study the generation of $V$ modes from photon-photon and photon-spin-1-massive-particle forward scatterings without relying on a specific interaction, which thus represent possible new signatures of physics beyond the Standard Model.
\end{abstract}


\maketitle 


\section{Introduction}

According to the standard lore, the cosmic microwave background (CMB) can only possess some amount of linear polarization (the so-called $Q$ and $U$ modes) \cite{Kosowsky:1994cy, Seljak:1996is,Zaldarriaga:1996xe, Kamionkowski:1996ks, Kamionkowski:1996zd, Hu:1997hp, Dodelson:2003, dodelson:2017}.  This is the result of the Compton scattering between CMB photons and electrons, and the gravitational redshift induced by cosmological perturbations of the metric. Instead, the CMB circular polarization (the so-called $V$ modes) is usually not considered, because the electron-photon Compton scattering cannot generate it at the classical level. On the other hand, several recent papers have proposed different theoretical mechanisms able to produce some amount of the $V$ modes~\cite{Kosowsky:1996yc,Colladay:1998,Giovannini:2002,Cooray:2002nm,Scoccola:2004ke,Campanelli:2004pm,Giovannini:2008,Alexander:2009,Finelli:2009,Zarei:2010,Motie:2012,Sawyer:2012gn,Xue:2014,Mohammadi:2013,Mohammadi:2014,De:2015, Ejlli:2016, Ejlli:2017,Alexander:2017bxe,Sadegh:2017rnr,Vahedi:2018,Alexander:2018iwy,Inomata:2018vbu,Marc:2018,Kamionkowski:2018,Montero-Camacho:2018vgs,Bartolo:2018igk,Bartolo:2019eac}. All these studies are motivated by the fact that, from the observational point of view, CMB $V$ modes are currently not excluded, as methods to improve the sensitivity of CMB experiments to circular polarization are underway~\cite{King:2016exc}. For instance, the SPIDER Collaboration has provided upper bounds on the power spectrum of circular polarization  $\ell(\ell + 1) C_{\ell}^{VV} /(2\pi)$ that are reported in a range from 141 $ \mu {\rm K}^2$ to 255 $ \mu {\rm K}^2$ at angular scales  $33< \ell <307 $ \cite{Nagy:2017csq}. More recently, the CLASS experiment improved these constraints in a range from 0.4 $ \mu {\rm K}^2$ to 13.5 $ \mu {\rm K}^2$ at angular scales  $1 < \ell <120$ \cite{Padilla:2019dhz}. These constraints are, in general, several orders of magnitude higher than the expected number of CMB $V$ modes predicted by most of the theoretical models that can be found in the literature (see Ref.~\cite{King:2016exc} for future detection prospects). 

The current authors have carried out a systematic study of the generation of the $V$-mode polarization by the forward scattering of CMB photons from spin-2 \cite{Bartolo:2018igk} and spin-1/2 \cite{Bartolo:2019eac} particles. In this paper, we continue our study of $V$-mode generation by examining the forward scattering of CMB photons with spin-1 particles. To keep our study as general as possible, we will initially not make any assumption about the nature of spin-1 particles or the kind of interactions. However, we will consider these particles as photons and the interactions to be Euler-Heisenberg interactions wherever we want to compare with previous literature. Moreover, we will work in the so-called \textit{quantum Boltzmann equation} (QBE) formalism. (See Refs.  \cite{Kosowsky:1994cy, Alexander:2009,Zarei:2010,Mohammadi:2013,Mohammadi:2014,Mohammadi:2015,Batebi:2016ocb,Sadegh:2017rnr, Shakeri:2018} for examples of applications of this formalism in the CMB context.)
 
The effects of the photon-photon forward scattering on the CMB polarization have been previously studied in Refs. \cite{Sawyer:2012gn,Sadegh:2017rnr,Inomata:2018vbu,Montero-Camacho:2018vgs}. The fundamental result is the production of CMB $V$ modes for \textit{Faraday conversion} of CMB linear polarization. However, there is an apparent contradiction between the different papers. For instance, Ref. \cite{Sadegh:2017rnr}, working within the QBE formalism, predicts an amount of $V$-mode signal that is much larger than what is predicted by Refs. \cite{Inomata:2018vbu,Montero-Camacho:2018vgs}, which work with a geometrical formalism focusing on the birefringence in the index of refraction of the cosmological medium. In this paper, we will show that the two formalisms are fully consistent and give the same prediction for the number of $V$ modes produced, correcting the formulas and estimates of Ref. \cite{Sadegh:2017rnr}.

The paper is organized as follows: In Sec. \ref{sec:2}, we will provide an introduction on how to use the quantum Boltzmann equation formalism to study the effects of the photon-photon forward scattering on CMB polarization. In Sec. \ref{sec:3}, we will study the  CMB polarization mixing induced by the photon-photon forward scattering mediated by a generic interaction. In Sec. \ref{sec:4}, we will derive the expected power-spectrum statistics of CMB $V$ modes provided by the photon-photon forward scattering through Euler-Heisenberg interactions. Moreover, we will provide the expected number of $V$ modes for a generic interaction, as a function of free parameters. In Sec. \ref{sec:5}, we will investigate the CMB polarization mixing induced by a hypothetical photon-spin-1-massive-particle forward scattering. Finally,  Sec. \ref{sec:6} contains our main conclusions.


\section{Photon-Photon forward scattering from QBE Formalism} \label{sec:2}

\subsection{Description of the formalism}

We start our analysis by introducing the formalism adopted for the rest of this paper. We parametrize the intensity and the polarization of CMB radiation through a density matrix $\rho_{ij}$, defined in terms of four Stokes parameters, in the following form \cite{Kosowsky:1994cy}:
\beq
 \rho_{ij}=\frac{1}{2}\left(
                   \begin{array}{cc}
                     I+Q & U-i V \\
                     U+i V & I-Q \\
                   \end{array}
                 \right)\, ,\label{eq:sb}
\eeq
where the parameter $I$ defines the intensity of unpolarized CMB radiation, $Q$ and $U$ define the CMB linear polarization, and $V$ refers to CMB circular polarization. The equations of motion for the Stokes parameters can be found through the so-called quantum Boltzmann equation, which is given in the literature as \cite{Kosowsky:1994cy}
\begin{align}
(2\pi)^3 \delta^{(3)}(0)(2k^0)
\frac{d\rho_{ij}(\mathbf k)}{dt} = i\left\langle \left[H_I
(0),\mathcal{D}_{ij}(\mathbf k)\right]\right\rangle-\frac{1}{2}\int_{-\infty}^{\infty} dt\left\langle
\left[H_I(t),\left[H_I
(0), \mathcal{D}_{ij}(\mathbf k)\right]\right]\right\rangle \, ,  \label{eq:Boltzmann equation}
\end{align}
where $k^0$ is the energy of CMB photons, $H_I(t)$ is the (effective) interaction Hamiltonian (describing in our case, e.g., the photon-photon interactions), and $\mathcal{D}_{ij}(\mathbf k) = a^\dagger_i(\mathbf k) a_j(\mathbf k)$ is the photon number operator [$a^\dagger_i(\mathbf k)$ and  $a_j(\mathbf k)$ being the creation and annihilation operators; see more details later]. Within this formalism, the expectation value of a generic operator $A$ is defined as \cite{Kosowsky:1994cy}
  \beq
  \langle A(\mathbf{k}) \rangle =\textrm{tr}[\rho \, A(\mathbf{k})]=\int\frac{d^3p}{(2\pi)^3} \langle \mathbf{p}| \rho \, A(\mathbf{k})|\mathbf{p} \rangle \, , \label{eq:expectation}
  \eeq
where $\rho$ denotes the following \textit{density operator}:
\beq
\rho = \int \frac{d^3 p}{(2 \pi)^3} \, \rho_{ij}(\mathbf p) \, \mathcal{D}_{ij}(\mathbf p) \, .
\eeq
In Eq. \eqref{eq:Boltzmann equation} the first term on the right-hand side is the so-called \textit{forward-scattering} term and the second term is the so-called \textit{damping} term. In this work, we will focus on the forward-scattering term, which is able to generate couplings between different polarization states.\footnote{This is the same physical mechanism that induces the resonance enhancement of neutrino oscillations in matter. See, e.g., Ref. \cite{Sigl:1992fn}.} In fact, Eq. \eqref{eq:Boltzmann equation} is derived by adopting a perturbative approach so that increasing powers of the interaction Hamiltonian $H_I(t)$ reduce the strength of the corresponding term. For this reason, in any fundamental interaction in the perturbative regime in which the forward scattering term is nonzero, \textit{a priori} it is expected to give the relevant physical effects on the CMB polarizations. Of course, this is not the case for the standard QED interaction between photons and electrons, where such a forward-scattering term vanishes (see, e.g., Ref. \cite{Kosowsky:1994cy}), and all the relevant effects arise from the damping term only. 

In particular, we are interested in the effects of the forward scattering of CMB photons with other (massless) spin-1 particles. Given $S^{(4)}$ as the $S$-matrix element describing this process, the (effective) interaction Hamiltonian can be defined through \cite{Karplus:1950}
\beq
S^{(4)}(\gamma(p_1)+\gamma(p_2)\rightarrow \gamma(p_3)+\gamma(p_4))=  i\int_{-\infty}^{\infty} dt H_{I}\, ,
\eeq
where $H_I$ can generally be written as
\beq
H_I= \int d\mathbf{p}^1 d\mathbf{p}^2 d\mathbf{p}^3 d\mathbf{p}^4 (2\pi)^{3}\delta^{(3)}(\mathbf{p}^3+\mathbf{p}^4-\mathbf{p}^1-\mathbf{p}^2) \times 3 \mathcal{M}(\mathbf{p}^1,r;\mathbf{p}^2,s;\mathbf{p}^3,r';\mathbf{p}^4,s')\, a^{\dag}_{r'}(p^3)a_{r}(p^1)a^{\dag}_{s'}(p^4) a_{s}(p^2)\, ,\label{eq:HI}
\eeq
where 
\beq
d\mathbf{p}\equiv \int \frac{d^3\mathbf{p}}{(2\pi)^3 2p^0} \, ,
\eeq
and $\mathcal{M}(\mathbf{p}^1,r;\mathbf{p}^2,s;\mathbf{p}^3,r';\mathbf{p}^4,s')$ is the Lorentz-invariant amplitude of this interaction, as a function of photon momenta and photon polarization indices $r,r',s,s'=1,2$, where $1$ and $2$ here stand for the two independent transverse polarizations. Moreover, $a_{i}(p)$ and $a_{i}^{\dagger}(p)$ denote the annihilation and creation operators, respectively, for photons obeying the following canonical commutation relation:
\beq
\left[a_s(p),a^{\dagger}_{s'}(p')\right]= (2 \pi)^3 ~ 2p^0 ~\delta^{(3)} (\mathbf{p}-\mathbf{p'})~ \delta_{s,s'}\label{eq:commutation} \, .
\eeq
Inserting Eq. \eqref{eq:HI} into the forward-scattering term of Eq. \eqref{eq:Boltzmann equation}, we get
\begin{align}                 
\left<\left[H_I
(0),\mathcal{D}_{ij}(\mathbf{k})\right]\right>=&\int d\mathbf{p}^1 d\mathbf{p}^2 d\mathbf{p}^3 d\mathbf{p}^4 (2\pi)^{3}\delta^{(3)}(\mathbf{p}^3+\mathbf{p}^4-\mathbf{p}^1-\mathbf{p}^2)~
\times 3 \mathcal{M}(\mathbf{p}^1,r;\mathbf{p}^2,s;\mathbf{p}^3,r';\mathbf{p}^4,s')\nonumber\\
& \times \left< a^{\dag}_{r'}(p^3)a_{r}(p^1)a^{\dag}_{s'}(p^4) a_{s}(p^2)a^{\dag}_{i}(k) a_{j}(k)
-a^{\dag}_{i}(k) a_{j}(k)a^{\dag}_{r'}(p^3)a_{r}(p^1)a^{\dag}_{s'}(p^4) a_{s}(p^2) \right>  \label{eq:commut}\, .
\end{align}
Using Eqs. \eqref{eq:expectation} and \eqref{eq:commutation}, we obtain the following expectation value of the product of photon creation and annihilation operators:
\beq \label{eq:contr_Wick}
\left<a^{\dag}_{m}(p)a_{n}(p')\right>=(2\pi)^3 ~ 2p^0 ~ \delta^{(3)}(\mathbf{p}-\mathbf{p}')\rho_{mn}(\mathbf{p})\, .
\eeq
Thus, using Eq. \eqref{eq:contr_Wick}, we can perform the expectation value in Eq. \eqref{eq:commut} by employing Wick's theorem, and, after integrating out three of the momenta with the Dirac deltas, we find the following final form of our Boltzmann equation\footnote{In this equation, $\rho^{\gamma}_{ij}$ refers to the density matrix of the observed CMB photons, while $\rho^{b}_{ij}$ denotes the density matrix of the ``background" target CMB photons.}:
\begin{align}
\frac{d\rho^{\gamma}_{ij}(\mathbf{k})}{dt}=& \frac{3i}{2 k^0} \int d\mathbf{p}~ \left(  \left[\delta_{is}\delta_{rs'} \rho^{\gamma}_{r'j}(\mathbf{k}) - \delta_{jr'}\delta_{rs'} \rho^{\gamma}_{is}(\mathbf{k}) + \delta_{jr'} \rho^{\gamma}_{is}(\mathbf{k}) \rho^{b}_{s'r}(\mathbf{p})  - \delta_{is}~ \rho^{\gamma}_{r'j}(\mathbf{k}) \rho^{b}_{s'r}(\mathbf{p}) \right] \mathcal{M}(\mathbf{p},r;\mathbf{k},s;\mathbf{k},r';\mathbf{p},s') \right. \nonumber\\
&\left. +\left[
   \delta_{is}  \rho^{\gamma}_{s'j}(\mathbf{k}) \rho^{b}_{r'r}(\mathbf{p})- \delta_{s' j}  \rho^{\gamma}_{is}(\mathbf{k})  \rho^{b}_{r'r}(\mathbf{p})\right]  \mathcal{ M}(\mathbf{p},r;\mathbf{k},s;\mathbf{p},r';\mathbf{k},s') \right. \nonumber\\
& \left. +\left[\delta_{ir} \rho^{\gamma}_{s' j}(\mathbf{k}) \rho^{b}_{r's}(\mathbf{p}) -  \delta_{js'} \rho^{\gamma}_{ir}(\mathbf{k}) \rho^{b}_{r's}(\mathbf{p})\right]  \mathcal{M}(\mathbf{k},r;\mathbf{p},s;\mathbf{p},r';\mathbf{k},s') \right. \nonumber\\
& \left.  +\left[\delta_{ir} \rho^{\gamma}_{r' j}(\mathbf{k}) \rho^{b}_{s's}(\mathbf{p}) -  \delta_{jr'} \rho^{\gamma}_{ir}(\mathbf{k}) \rho^{b}_{s's}(\mathbf{p})\right]   \mathcal{M}(\mathbf{k},r;\mathbf{p},s;\mathbf{k},r';\mathbf{p},s')  \right) \, , \label{eq:evolution of rho}
\end{align}
where $ \mathbf{p}$ and $ \mathbf{k}$ indicate the momenta of the background ($b$) and of the line-of-sight observed ($\gamma$) photons respectively. In the next section, we will employ the latter equation to evaluate the effects of photon-photon forward scattering mediated by Euler-Heisenberg interactions on the CMB polarization field. However, before doing this, in the following subsections we will introduce a general parametrization of the photon-photon scattering amplitude.


\subsection{General photon-photon scattering amplitude}

\begin{figure}
   \includegraphics[width=2.5in]{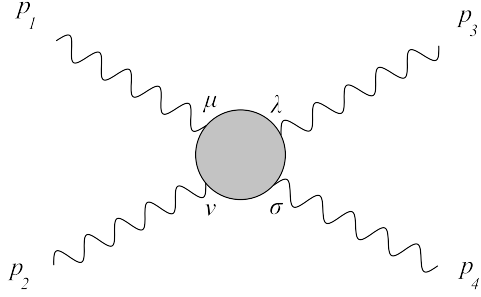}\\
  \caption{An example of a Feynman diagram associated with photon-photon scattering.}\label{fig:feyndiag}
\end{figure}

In this subsection, by using symmetry considerations, we introduce a general amplitude describing the scattering of two massless spin-1 particles which does not rely on any specific photon-photon fundamental interaction. In fact, assuming that we work in the context of a quantum field theory (QFT) that is unitary and where all the interactions are local, we can employ the following general parametrization for the photon-photon Lorentz-invariant scattering amplitude \cite{Karplus:1950, Leo:1975, Costantini:1971}:
\beq
\mathcal M=M_{\mu \nu \lambda \sigma}(1234)\, \epsilon^1_{\mu}  \epsilon^2_{\nu} \epsilon^3_{\lambda}  \epsilon^4_{\sigma} \, ,\label{eq:general_amplitude}
\eeq
where  $\epsilon^i_{\mu}\equiv \epsilon_{\mu}(p^{i})$ are the polarization vectors of incoming and outgoing photons, and $M_{\mu \nu \lambda \sigma}(1234)\equiv M_{\mu \nu \lambda \sigma}(p^1 , p^2, p^3, p^4)$, with $p^1$ and $p^2$ ($p^3$ and $p^4$) denoting the four-momenta of incoming (outgoing) photons in the $\gamma(p_1)+\gamma(p_2)\rightarrow \gamma(p_3)+\gamma(p_4)$ scattering process. The four-rank tensor $M_{\mu \nu \lambda \sigma}(1234)$ must respect the crossing and gauge symmetries. The gauge symmetry implies the following identities:
\beq
 p_{1}^{\mu}M_{\mu \nu \lambda \sigma}=0 \, ,  \qquad \qquad p_{2}^{\nu}M_{\mu \nu \lambda \sigma}=0\, , \qquad  \qquad p_{3}^{\lambda}M_{\mu \nu \lambda \sigma}=0\, , \qquad \qquad p_{4}^{\sigma}M_{\mu \nu \lambda \sigma}=0 \,  ,
\eeq
while, due to the crossing symmetry, $M_{\mu \nu \lambda \sigma}(p^1 , p^2, p^3, p^4)$ is given by summing over all the $4!$ possible permutations of external photons with momenta $p^1$, $p^2$, $p^3$, and $p^4$ and simultaneously their corresponding vertex indices. We have depicted one of these terms in Fig. \ref{fig:feyndiag}.

We can expand $M_{\mu \nu \lambda \sigma}(p^1 , p^2, p^3, p^4)$ in terms of a set of  four-rank independent tensors $T^{(i)}_{\mu \nu \lambda \sigma}$ as \cite{Leo:1975}
\beq
M_{\mu \nu \lambda \sigma}(1234)=\displaystyle\sum_{i=1}^{5}G_i(s, t, u) T^{(i)}_{\mu \nu \lambda \sigma}(1234) \, ,\label{eq:general tensor}
\eeq
where the coefficients $G_i(s, t, u)$ are invariant scalar amplitudes that may depend on invariant kinematics, as the Mandelstam variables 
\beq
s=(p^1+p^2)^2 \, , \qquad \qquad t=(p^1-p^3)^2 \, , \qquad \qquad u=(p^1-p^4)^2 \, .
\eeq
Moreover, the tensors $T^{(i)}_{\mu \nu \lambda \sigma}$ can be expressed in terms of a tensor basis $f^{(i)}_{\mu \nu \lambda \sigma}$  as \cite{Karplus:1950, Leo:1975}
\begin{align}
T^{(1)}_{\mu \nu \lambda \sigma}(1234)=& f^{(1)}_{\mu \nu \lambda \sigma}(1234)\, , \nonumber \\
T^{(2)}_{\mu \nu \lambda \sigma}(1234)=& f^{(1)}_{\lambda \nu \mu \sigma}(3214)\, ,\nonumber \\
T^{(3)}_{\mu \nu \lambda \sigma}(1234)=& f^{(1)}_{\sigma \nu \lambda \mu}(4231)\, , \nonumber \\
T^{(4)}_{\mu \nu \lambda \sigma}(1234)=& f^{(2)}_{\mu \nu \lambda \sigma}(1234)+f^{(2)}_{\mu \nu  \sigma \lambda}(1243)+f^{(2)}_{\nu \lambda \mu  \sigma }(2314)\, , \nonumber \\
T^{(5)}_{\mu \nu \lambda \sigma}(1234)=& f^{(3)}_{\mu \nu \lambda \sigma}(1234)+f^{(3)}_{\nu \mu  \sigma \lambda }(2143)+f^{(3)}_{ \lambda \sigma\mu \nu}(3412)+f^{(3)}_{ \sigma \lambda\nu\mu }(4312)\nonumber\\
&+f^{(3)}_{\mu \lambda\nu  \sigma}(1324)+f^{(3)}_{\lambda\mu \sigma\nu  }(3142)+f^{(3)}_{ \nu \sigma\mu\lambda }(2413)+f^{(3)}_{ \sigma\nu \lambda\mu }(4231)\nonumber\\
& +f^{(3)}_{\mu \sigma\lambda\nu  }(1432)+f^{(3)}_{\sigma\mu \nu \lambda }(4123)+f^{(3)}_{ \lambda\nu \mu \sigma}(3214)+f^{(3)}_{ \nu \lambda \sigma\mu}(2341)\, ,
\end{align}
where the tensor basis is defined in the following equations
\begin{align}
 f^{(1)}_{\mu \nu \lambda \sigma}(1234) =&p^2_{\mu} p^1_{\nu} p^4_{\lambda} p^3_{\sigma} - (p^3 \cdot p^4)g_{\lambda \sigma} p^2_{\mu} p^1_{\nu}- (p^1 \cdot p^2)g_{\mu \nu} p^4_{\lambda} p^3_{\sigma}+ (p^1 \cdot p^2) (p^3 \cdot p^4) g_{\mu \nu} g_{\lambda \sigma} \label{eq:f1}\, ,\\
 f^{(2)}_{\mu \nu \lambda \sigma}(1234)=& p^2_{\mu}p^3_{\nu} p^4_{\lambda} p^1_{\sigma}+ p^4_{\mu} p^1_{\nu} p^2_{\lambda} p^3_{\sigma} -(p^1 \cdot p^4)g_{\lambda \sigma} p^2_{\mu} p^3_{\nu} - (p^3 \cdot p^4)g_{\nu \lambda } p^2_{\mu} p^1_{\sigma}+(p^1\cdot p^4)g_{\nu \lambda } p^2_{\mu} p^3_{\sigma} - (p^2 \cdot p^3)g_{\mu \nu} p^4_{\lambda} p^1_{\sigma}\nonumber\\
&+ (p^3 \cdot p^4)g_{\mu \nu} p^2_{\lambda} p^1_{\sigma}- (p^1 \cdot p^4)g_{\mu \nu} p^2_{\lambda} p^3_{\sigma} - (p^1 \cdot p^2)g_{\mu \sigma} p^3_{\nu} p^4_{\lambda}+(p^1 \cdot p^2)g_{\lambda \sigma} p^3_{\nu} p^4_{\mu} - (p^1\cdot p^2)g_{\nu \lambda} p^4_{\mu} p^3_{\sigma}\nonumber\\
&+(p^2 \cdot p^3)g_{\mu \sigma} p^1_{\nu} p^4_{\lambda} - (p^2 \cdot p^3)g_{\lambda \sigma} p^4_{\mu} p^1_{\nu} - (p^3 \cdot p^4)g_{\mu \sigma} p^1_{\nu} p^2_{\lambda}+(p^1 \cdot p^4)(p^2 \cdot p^3) g_{\mu \nu} g_{\lambda \sigma}\nonumber\\
&+(p^1 \cdot p^2)(p^3 \cdot p^4) g_{\mu \sigma} g_{\nu \lambda } \label{eq:f2} \, , \\
 f^{(3)}_{\mu \nu \lambda \sigma}(1234) =&(p^3 \cdot p^4) p^2_{\mu}p^1_{\nu} p^1_{\lambda} p^1_{\sigma} -(p^1 \cdot p^3) p^2_{\mu}p^1_{\nu} p^4_{\lambda} p^1_{\sigma} - (p^1 \cdot p^4) p^2_{\mu}p^1_{\nu} p^1_{\lambda} p^3_{\sigma}+ (p^1 \cdot p^3)(p^1 \cdot p^4) g_{\lambda \sigma} p^2_{\mu}p^1_{\nu}\nonumber\\
& + (p^1 \cdot p^2)(p^1 \cdot p^3) g_{\mu \nu} p^4_{\lambda}p^1_{\sigma} -  (p^1 \cdot p^2)(p^3 \cdot p^4) g_{\mu \nu} p^1_{\lambda}p^1_{\sigma}+ (p^1 \cdot p^2)(p^1 \cdot p^4) g_{\mu \nu} p^1_{\lambda}p^3_{\sigma}\nonumber\\
& - (p^1 \cdot p^2) (p^1 \cdot p^3)(p^1 \cdot p^4) g_{\mu \nu} g_{\lambda \sigma}  \label{eq:f3} \, .
\end{align}
Thus, using Eq.  \eqref{eq:general_amplitude}, we can express the photon-photon scattering amplitude as a function of the metric tensor $g_{\mu \nu}$, the photon four-momenta and polarization vectors, and generic coefficients without specifying a given fundamental interaction.


\subsection{QED case: Euler-Heisenberg amplitude}

\begin{figure}
   \includegraphics[width=5.5in]{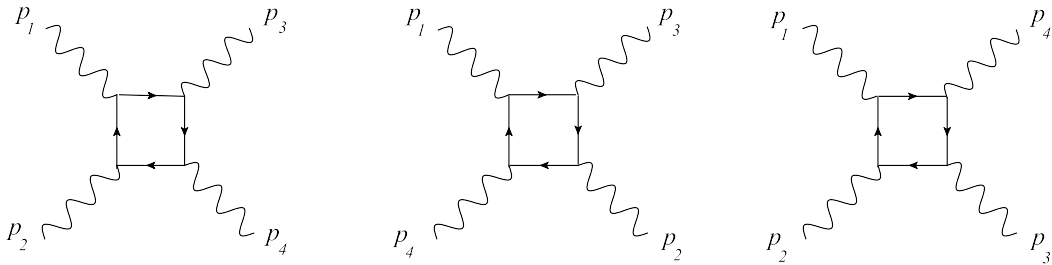}
  \caption{Three independent one-loop Feynman diagrams describing the photon-photon scattering for clockwise electron loop direction. In all the diagrams, $p_1$ and $p_2$ denote the incoming photons' momenta, while $p_3$ and $p_4$ refer to the outgoing momenta \cite{Berestetsky:1982aq}. }\label{fig:feyndiag1}
\end{figure}

In the quantum electrodynamics (QED) context, photon-photon interactions are described by the so-called Euler-Heisenberg Lagrangian, which is a low-energy effective Lagrangian describing multiple photon interactions.~This~reads~\cite{Euler:1935zz, Heisenberg:1935qt, Karplus:1950}
\beq
\mathcal{L}_{\rm{E - H}} = \alpha_1  \left(F_{\mu \nu}(x) F_{\mu \nu}(x)\right)^2 +  \alpha_2\left(F_{\alpha \beta}(x)  F^{\beta \gamma}(x)  F_{\gamma \rho}(x)    F^{\rho \alpha}(x)\right) \, ,
 \label{eq:low energy1}
\eeq
 where
\beq
\alpha_1 =\frac{5\alpha^2}{180 m_e^4}~,\:\:\:\:\:\:\:\: \textrm{and}\:\:\:\:\:\:\:\:  \alpha_2 =-\frac{14\alpha^2}{180 m_e^4}\, ,
\eeq
where $\alpha=e^2/(4 \pi)$ denotes the so-called fine-structure constant, $m_e$ is the electron mass and $F_{\mu \nu} = \partial_\mu A_\nu - \partial_\nu A_\mu $ is the well-known photon field strength. This Lagrangian can be also expressed in terms of the electric and magnetic fields as
\beq
\mathcal{L}_{\rm{E - H}} =  a (\mathbf{E}^2 - \mathbf{B}^2)^2 +b(\mathbf{E} \cdot \mathbf{B})^2\, ,
 \label{eq:low_energy1}
\eeq
where 
\beq
a=\frac{2\alpha^2}{45 m_e^4}~,\:\:\:\:\:\:\:\: \textrm{and}\:\:\:\:\:\:\:\:  b=\frac{14\alpha^2}{45m_e^4}\, .
\eeq
This effective Lagrangian is derived by the photon-photon scattering process mediated by the one-loop box Feynman diagrams containing electrons in the internal lines (see, e.g., Fig. \ref{fig:feyndiag1}) in the low-energy limit where the external photons are soft, with energies much lower than the electron mass $m_e$. It is possible to show that the Feynman amplitude derived by the Euler-Heisenberg Lagrangian can be expressed  in terms of the basis tensors $f^{(i)}_{\mu \nu \lambda \sigma}$ as \cite{Karplus:1950} 
\begin{align}
M_{\mu \nu \lambda \sigma} (1234)=&\frac{4 \alpha^2}{9 m_e^4 } \left(f^{(1)}_{\mu \nu \lambda \sigma}(1234)+f^{(1)}_{\mu  \lambda\nu \sigma}(1324)+ f^{(1)}_{\mu \sigma\nu \lambda }(1423)\right) \nonumber \\
& - \frac{14 \alpha^2}{45 m_e^4 } \left(f^{(2)}_{\mu \nu \lambda \sigma}(1234)+ f^{(2)}_{\mu  \lambda\nu \sigma}(1324)+ f^{(2)}_{\mu \sigma\nu \lambda }(1423)\right) \,, \label{eq:Euler-Heisenberg}
\end{align}
which is included in the general form in Eq. \eqref{eq:general tensor}.

\section{Polarization mixing from photon-photon forward scattering} \label{sec:3}

Now, the effect of the photon-photon forward scattering on the dynamics of CMB Stokes parameters is obtained by inserting Eq. \eqref{eq:general_amplitude} into the Boltzmann equation [Eq. \eqref{eq:evolution of rho}]. Because we are finally interested in the polarization and intensity of CMB radiation, we first give the expression of the Stokes parameters in terms of the CMB density matrix. In this respect, the unperturbed CMB photon density matrix is written as \cite{Kosowsky:1994cy}
\beq
 \rho^{(0)}_{ij}(k)=\frac{1}{2}\left(
                   \begin{array}{cc}
                     I_0(k) & 0 \\\\
                     0 & I_0(k) \\
                   \end{array}
                 \right)\, ,\label{eq:density0}
\eeq
while the CMB radiation field perturbations are defined as \cite{Kosowsky:1994cy}~\footnote{These are analogous to the brightness perturbations defined in  Ref. \cite{Kosowsky:1994cy} apart from a factor $4$---i.e., our $\Delta_I^\gamma$ is a factor $(1/4)$ of the $\Delta_I$ quantities defined in Ref.~\cite{Kosowsky:1994cy} [see, e.g., Eq. (6.51) of Ref.~\cite{Kosowsky:1994cy}]---so that, e.g., here $\Delta_I^\gamma$ represents the temperature fluctuations [it would correspond to the quantity $\Theta$ defined, e.g., in Ref. ~\cite{Dodelson:2003} or in Eq. (5.3) of Ref.~\cite{Bartolo:2006cu}. For a discussion of the various temperature variables that can be used, see, e.g., Refs.~\cite{Bartolo:2011wb,Pitrou:2010sn}].}
\beq
\left[k^0_c\frac{\partial I_0(k_c)}{\partial k^0_c}\right]^{-1} \rho_{ij}^{(1)}(\mathbf{x,k}_c)=\frac{1}{2}\left(
                   \begin{array}{cc}
                     \Delta^{\gamma}_I(\mathbf{x},{\mathbf{ k}}_c)+\Delta^{\gamma}_Q(\mathbf{x},{\mathbf{ k}}_c) ~&~ \Delta^{\gamma}_U(\mathbf{x},{\mathbf{ k}}_c)-i\Delta^{\gamma}_V(\mathbf{x},{\mathbf{ k}}_c) \\\\\\
                     \Delta^{\gamma}_U(\mathbf{x},{\mathbf{ k}}_c)+i\Delta^{\gamma}_V(\mathbf{x},{\mathbf{ k}}_c) ~&~ \Delta^{\gamma}_I(\mathbf{x},{\mathbf{ k}}_c)-\Delta^{\gamma}_Q(\mathbf{x},{\mathbf{ k}}_c) \\
                   \end{array}
                 \right)\, ,\label{eq:density1}
\eeq
where $k_c=a k$ is the comoving wave number of CMB photons, with $a(\eta)$ denoting the scale factor as a function of conformal time $d \eta=dt/a(t)$, and $I_0(k)=(e^{k/T}-1)^{-1}$ is the Bose-Einstein distribution function describing the homogeneous (unperturbed) distribution of CMB photons. As above, the upper index $\gamma$ refers to the observed photons. In the same way, the background beam is described by
\beq
 \rho_{ij}^{(0)}(p)=\frac{1}{2}\left(
                   \begin{array}{cc}
                     I_0(p) & 0 \\\\
                     0 & I_0(p) \\
                   \end{array}
                 \right)\, ,\label{eq:density2}
\eeq
 and
\beq
\left[p^0 \frac{\partial I_0(p)}{\partial p^0}\right]^{-1}   \rho_{ij}^{(1)}(\mathbf{x},\mathbf{p})=\frac{1}{2}\left(
                   \begin{array}{cc}
                     I^b(\mathbf{x},\hat{\mathbf{ p}})+Q^b(\mathbf{x},\hat{\mathbf{ p}}) ~&~ U^b(\mathbf{x},\hat{\mathbf{ p}})-iV^b(\mathbf{x},\hat{\mathbf{ p}})\\\\\\
                     U^b(\mathbf{x},\hat{\mathbf{ p}})+iV^b(\mathbf{x},\hat{\mathbf{ p}})~&~ I^b(\mathbf{x},\hat{\mathbf{ p}})-Q^b(\mathbf{x},\hat{\mathbf{ p}}) \\
                   \end{array}
                 \right)\, ,\label{eq:density3}
\eeq
where this time the upper index $b$ refers to the background photons. [We use slightly different notations with respect to Eq.~(\ref{eq:density1}) to easily keep track of the background target beam.] 

Thus, using these definitions for the photon density matrices, we insert the general scattering amplitude [Eq. \eqref{eq:general_amplitude}\footnote{It is understood that we are using the FRW metric $ds^2 = -dt^2 + a^2 d\mathbf x^2$ in evaluating Eq. \eqref{eq:general_amplitude}.}] into the Boltzmann equation [Eq. \eqref{eq:evolution of rho}] and sum over all the vector polarization indices. After some straightforward calculations, we find
\begin{align}
\frac{d}{d\eta}{\Delta}^{\gamma}_I(\mathbf{x,{k}}_c)=0\, ,
\end{align}
which is expected, since there is no energy or momentum transfer in the forward scattering of photons, and
\begin{align}
\frac{d}{d\eta}{\Delta}^{\gamma}_Q(\mathbf{x},{\mathbf{ k}}_c)=& - \frac{3a^2(\eta)}{k^0_c} ~ \int \frac{d^3\mathbf{p}}{(2\pi)^3 2p^0}\left[p^0 \frac{\partial I_0(p)}{\partial p^0}\right] \left\{ \left[g_1~ I^b(\mathbf{x},\hat{\mathbf{ p}})+ g_2~ U^b(\mathbf{x},\hat{\mathbf{ p}})+g_3~ Q^b(\mathbf{x},\hat{\mathbf{ p}}) + s_1(\mathbf{p,k}_c)\right]\Delta_V^\gamma(\mathbf{x},\mathbf{{k}}_c)\right.\nonumber\\
&\:\:\:\:\:\:\:\:\:\:\:\: \:\:\:\:\:\: \:\:\:\:\:\: \:\:\:\:\:\: \:\:\:\:\:\: \:\:\:\:\:\: \:\:\:\:\:\:  \left. + g_4~ V^b(\mathbf{x},\hat{\mathbf{ p}}) \Delta_U^\gamma(\mathbf{x},{\mathbf{ k}}_c) \right\}\, , \label{eq:SPQ}\\
\frac{d}{d\eta}{\Delta}^{\gamma}_U(\mathbf{x,{k}}_c)=& \frac{3a^2(\eta)}{2 k^0_c} ~   \int \frac{d^3\mathbf{p}}{(2\pi)^3 2p^0} \left[p^0 \frac{\partial I_0(p)}{\partial p^0}\right] \left\{ \left[g_5~ I^b(\mathbf{x},\hat{\mathbf{ p}}) + g_6~ U^b(\mathbf{x},\hat{\mathbf{ p}}) +  g_7~ Q^b(\mathbf{x},\hat{\mathbf{ p}}) +s_2(\mathbf{p,k}_c)\right] \Delta_V^\gamma(\mathbf{x,{k}}_c)  \right. \nonumber\\
& \:\:\:\:\:\:\:\:\:\:\:\: \:\:\:\:\:\: \:\:\:\:\:\: \:\:\:\:\:\: \:\:\:\:\:\: \:\:\:\:\:\: \: \: \:\: \left.+2 g_4~ V^b(\mathbf{x},\hat{\mathbf{ p}})\Delta_Q^\gamma(\mathbf{x,{k}}_c)\right\}\, ,\label{eq:SPU} \\
\frac{d}{d\eta}{\Delta}^{\gamma}_V(\mathbf{x,{k}}_c)=& -\frac{3a^2(\eta)}{2 k^0_c}~  \int \frac{d^3\mathbf{p}}{(2\pi)^3 2p^0}\left[p^0 \frac{\partial I_0(p)}{\partial p^0}\right]  \left\{  \left[ g_5~ I^b(\mathbf{x},\hat{\mathbf{ p}}) + g_6~ U^b(\mathbf{x},\hat{\mathbf{ p}})+g_7~ Q^b(\mathbf{x},\hat{\mathbf{ p}})  + s_2(\mathbf{p,k_c})\right] \Delta_U^\gamma(\mathbf{x,{k}}_c)\right. \nonumber\\
& \left. -2  \left[g_1~ I^b(\mathbf{x},\hat{\mathbf{ p}}) +g_2~ U^b(\mathbf{x},\hat{\mathbf{ p}})+g_3~ Q^b(\mathbf{x},\hat{\mathbf{ p}})+s_1(\mathbf{p,k}_c)\right] \Delta_Q^\gamma(\mathbf{x,{k}}_c)   \right\} \, ,\label{eq:SPV}
\end{align}
where the $g_i$ coefficients and the scalar functions $s_i$  are given in Appendix \ref{appendix_A}.

From the physical point of view, the set of coupled Eqs.  \eqref{eq:SPQ}, \eqref{eq:SPU}, and \eqref{eq:SPV} just derived gives rise both to the transformation of $Q$ modes into $U$ modes and vice-versa (\textit{Faraday rotation}), and to the conversion of linear polarization to circular polarization and vice-versa (\textit{Faraday conversion}). In this paper, we are interested only in the Faraday conversion effect; thus, we decouple $Q$ and $U$ modes by assuming $g_4 = 0$ (or $G_1 = G_2$), leaving
\begin{align}
\frac{d}{d\eta}{\Delta}^{\gamma}_Q(\mathbf{x,{k}}_c)=& - \frac{ 3a^2(\eta)}{ k^0_c}  ~ \int \frac{d^3\mathbf{p}}{(2\pi)^3 2p^0}\left[p^0 \frac{\partial I_0(p)}{\partial p^0}\right]\left[g_1~ I^b(\mathbf{x},\hat{\mathbf{ p}})+ g_2~ U^b(\mathbf{x},\hat{\mathbf{ p}}) +g_3~ Q^b(\mathbf{x},\hat{\mathbf{ p}})  + s_1(\mathbf{p,k}_c)\right]  \Delta_V^\gamma(\mathbf{x,{k}}_c) \, , \label{eq:SPQ2}\\
\frac{d}{d\eta}{\Delta}^{\gamma}_U(\mathbf{x,{k}}_c)=& \frac{ 3a^2(\eta)}{2 k^0_c} ~  \int \frac{d^3\mathbf{p}}{(2\pi)^3 2p^0}\left[p^0 \frac{\partial I_0(p)}{\partial p^0}\right] \left[g_5~ I^b(\mathbf{x},\hat{\mathbf{ p}}) + g_6~ U^b(\mathbf{x},\hat{\mathbf{ p}}) +  g_7~ Q^b(\mathbf{x},\hat{\mathbf{ p}}) +s_2(\mathbf{p,k}_c)\right]  \Delta_V^\gamma(\mathbf{x,{k}}_c)  \, ,\label{eq:SPU2}\\
\frac{d}{d\eta}{\Delta}^{\gamma}_V(\mathbf{x,{k}}_c)=& -\frac{3a^2(\eta)}{2 k^0_c}  ~ \int \frac{d^3\mathbf{p}}{(2\pi)^3 2p^0}\left[p^0 \frac{\partial I_0(p)}{\partial p^0}\right] \left\{ \left[ g_5~ I^b(\mathbf{x},\hat{\mathbf{ p}}) + g_6~ U^b(\mathbf{x},\hat{\mathbf{ p}})+g_7~ Q^b(\mathbf{x},\hat{\mathbf{ p}})  + s_2(\mathbf{p,k_c})\right]  \Delta_U^\gamma(\mathbf{x,{k}}_c) \right. \nonumber\\
& \left.  -2  \left[g_1~ I^b(\mathbf{x},\hat{\mathbf{ p}})+g_2~ U^b(\mathbf{x},\hat{\mathbf{ p}})+g_3~ Q^b(\mathbf{x},\hat{\mathbf{ p}})+s_1(\mathbf{p,k}_c)\right] \Delta_Q^\gamma(\mathbf{x,{k}}_c)  \right\} \, .\label{eq:SPV2}
\end{align}


\subsection{Euler-Heisenberg case}


In this subsection, we derive in our quantum Boltzmann equation formalism the linear-circular polarization mixing induced by Euler-Heisenberg interactions. Thus, substituting the Euler-Heisenberg Feynman amplitude \eqref{eq:Euler-Heisenberg} into Eq.~\eqref{eq:evolution of rho}, we get the following set of equations:
\begin{align}
\frac{d}{d\eta}{\Delta}^\gamma_I(\mathbf{x, k}_c)  =&\, 0 \, , \\
\frac{d}{d\eta}{\Delta}_Q^\gamma(\mathbf{x, k}_c)=& - \frac{4\alpha ^2 a^2(\eta)}{15 k^0_c m_e^4}   \Delta_V^\gamma(\mathbf{x,k}_c)  \int \frac{d^3\mathbf{p}}{(2\pi)^3 2p^0} \left[p^0 \frac{\partial I_0(p)}{\partial p^0}\right] \left[f_1 ~ I^b(\mathbf{x,\hat{p}}) + f_2~ U^b(\mathbf{x,\hat{p}})+f_3~ Q^b(\mathbf{x,\hat{p}}) + f_1 \right] \label{eq:SSSPQ} \, , \\
\frac{d}{d\eta}{\Delta}_U^\gamma(\mathbf{x, k}_c)=& \frac{2\alpha^2 a^2(\eta)}{15 k^0_c m_e^4}   \Delta_V^\gamma(\mathbf{x, k}_c)  \int \frac{d^3\mathbf{p}}{(2\pi)^3 2p^0} \left[p^0 \frac{\partial I_0(p)}{\partial p^0}\right] \left[f_4 ~I^b(\mathbf{x,\hat{p}})+ f_5~U^b(\mathbf{x,\hat{p}})+f_6~Q^b(\mathbf{x,\hat{p}}) +  f_3 \right]\label{eq:SSSPU} \, , \\
\frac{d}{d\eta}{\Delta}_V^\gamma(\mathbf{x, k}_c)=&- \frac{2\alpha^2 a^2(\eta)}{15 k^0_c m_e^4}    \int \frac{d^3\mathbf{p}}{(2\pi)^3 2p^0} \left[p^0 \frac{\partial I_0(p)}{\partial p^0}\right] \left\{\Delta_U^\gamma(\mathbf{x, k}_c) \left[f_4~I^b(\mathbf{x,\hat{p}})+ f_5~U^b(\mathbf{x,\hat{p}})+ f_6~Q^b(\mathbf{x,\hat{p}})  +  f_3 \right]\right. \nonumber\\
& \left. -2 \Delta_Q^\gamma(\mathbf{x, k}_c)\left[f_1~I^b(\mathbf{x,\hat{p}})+ f_2~U^b(\mathbf{x,\hat{p}})+ f_3~Q^b(\mathbf{x,\hat{p}})  +  f_1 \right]\right\}\, ,\label{eq:SSSPV}
\end{align}
where the explicit expressions for the $f_i$ coefficients are given in Appendix \ref{appendix_B}. Notice that these equations can be directly derived by Eqs. \eqref{eq:SPQ2}, \eqref{eq:SPU2}, and \eqref{eq:SPV2} once we identify
\beq \label{eq:EH_to_general}
G_1+G_2+2G_4 = \frac{4\alpha^2}{15 m_e^4} \, .
\eeq 
Moreover, notice that by matching the amplitudes \eqref{eq:general tensor} and \eqref{eq:Euler-Heisenberg}, we get
\beq
G_1 = G_2 \, ,
\eeq 
telling us that photon-photon scattering, as predicted by QED, leads only to Faraday conversion (moreover, in the low-energy limit, one also finds $G_3=0$; see Ref.~\cite{Leo:1975}). 

Now, in order to perform the integral over $\mathbf{p}$, we write the momenta and photon polarization vectors in the following general form:
\begin{align}
 \mathbf{\hat k}_c =& \,(\sin\theta\cos\phi,\,\sin\theta\sin\phi,\,\cos\theta)\, ,\nonumber\\
\hat{\mathbf{p}} =&\, (\sin\theta'\cos\phi',\,\sin\theta'\sin\phi',\,\cos\theta')\, ,\nonumber\\
 \bm{\epsilon}_{1}(k) =&\, (\cos\theta\cos\phi,\,\cos\theta\sin\phi,\,-\sin\theta)\, ,\nonumber\\
 \bm{\epsilon}_{2}(k) =& \,(-\sin\phi,\,\cos\phi,\,0)\, ,\nonumber\\
 \bm{\epsilon}_{1}(p) =& \,(\cos\theta'\cos\phi',\,\cos\theta'\sin\phi',\,-\sin\theta')\, ,\nonumber\\
 \bm{\epsilon}_{2}(p) =& \, (-\sin\phi',\,\cos\phi',\,0)\, .
\end{align}
In this generic reference frame, we get
\begin{align}
f_1 =&0\, ,\nonumber\\
f_2=& \frac{3}{8} (k^0 p^0)^2 \{-12 \sin^2\theta \sin^2\theta^\prime +16 \cos \theta \cos\theta' \cos2(\phi-\phi')-(\cos 2\theta+3)(\cos 2\theta'+3)\cos2\phi \cos 2\phi' \nonumber\\
&+4\cos (\phi-\phi')(4\sin \theta \sin \theta'-\sin 2\theta \sin 2\theta')- (\cos 2\theta+3)(\cos 2\theta'+3) \sin 2\phi \sin 2\phi'\}\, ,\nonumber\\
f_3=&6(k^0 p^0)^2 (\cos\theta-\cos\theta^\prime) \sin(\phi-\phi^\prime) \{(\cos\theta \cos\theta^\prime -1) [(\cos\phi \cos\phi^\prime+ \sin\phi \sin\phi^\prime]+\sin\theta \sin\theta^\prime \}\nonumber \, ,\\
f_4=&0\, , \nonumber\\
f_5=&12(k^0 p^0)^2 (\cos\theta-\cos\theta^\prime) \sin(\phi-\phi^\prime) \{(\cos\theta \cos\theta^\prime -1) [(\cos\phi \cos\phi^\prime+ \sin\phi \sin\phi^\prime]+\sin\theta \sin\theta^\prime \}\nonumber \, ,\\
f_6=& \frac{3}{4} (k^0 p^0)^2 \{-6 \cos2\theta  \sin^2\theta^\prime -16 \cos \theta \cos\theta' \cos2(\phi-\phi')+(\cos 2\theta+3)(\cos 2\theta'+3)\cos2\phi \cos 2\phi' \nonumber\\
&-3\cos2\theta^\prime-4\cos (\phi-\phi')(4\sin \theta \sin \theta'-\sin 2\theta \sin 2\theta')+ (\cos 2\theta+3)(\cos 2\theta'+3) \sin 2\phi \sin 2\phi'+3\}\, .
\end{align}
Now, without losing generality, we fix the frame where the line of sight is aligned with the $z$ axis---i.e., $\mathbf{\hat k}_c \parallel \hat z$---as in the end we will work with quantities that are invariant under rotations. Thus, we get
\begin{align}
f_2&= -3 (k^0 p^0)^2 \cos 2\phi' (1-\cos\theta')^2\, ,\nonumber\\
f_3&=3 (k^0 p^0)^2  \sin 2\phi' (1-\cos\theta')^2\, ,\nonumber\\
f_5&=6 (k^0 p^0)^2  \sin 2\phi' (1-\cos\theta')^2\, , \nonumber\\
f_6&=6 (k^0 p^0)^2  \cos 2\phi' (1-\cos\theta')^2\, . \label{eq:f-coeffs}
\end{align}
Hence, Eqs. \eqref{eq:SSSPQ}, \eqref{eq:SSSPU}, and \eqref{eq:SSSPV} become
\begin{align}
\frac{d}{d\eta}{\Delta}_Q^\gamma(\mathbf{x},k_c)&= - \frac{ 2\alpha ^2 k^0_c}{5  m_e^4}   \Delta_V^\gamma(\mathbf{x},k_c)  \int \frac{d^3\mathbf{p}}{(2\pi)^3 }\, \left[p^0 \frac{\partial I_0(p)}{\partial p^0}\right]\,p^0 (1-\cos\theta')^2 \left[   \sin 2\phi' ~ Q^b(\mathbf{x,\hat{p}}) -  \cos 2\phi' ~ U^b(\mathbf{x,\hat{p}})  \right] \label{eq:SSSSPQ} \, ,\\
\frac{d}{d\eta}{\Delta}_U^\gamma(\mathbf{x},k_c)&= \frac{2 \alpha ^2 k^0_c}{5 m_e^4}  \Delta_V^\gamma(\mathbf{x},k_c)  \int \frac{d^3\mathbf{p}}{(2\pi)^3 } \,\left[p^0 \frac{\partial I_0(p)}{\partial p^0}\right]\,p^0 (1-\cos\theta')^2 \left[ \cos 2\phi'~Q^b(\mathbf{x,\hat{p}}) + \sin 2\phi' ~U^b(\mathbf{x,\hat{p}})  \right]\label{eq:SSSSPU} \, , \\
\frac{d}{d\eta}{\Delta}_V^\gamma(\mathbf{x},k_c)&= - \frac{2\alpha ^2 k^0_c}{5  m_e^4}    \int \frac{d^3\mathbf{p}}{(2\pi)^3 } \,\left[p^0 \frac{\partial I_0(p)}{\partial p^0}\right]\, p^0 (1-\cos\theta')^2 \left\{   \Delta_U^\gamma(\mathbf{x},k_c) \left[   \cos 2\phi' ~Q^b(\mathbf{x,\hat{p}})  +\sin 2\phi' ~U^b(\mathbf{x,\hat{p}}) \right]\right. \nonumber\\
& \left. - \Delta_Q^\gamma(\mathbf{x},k_c)  \left[ \sin 2\phi'~Q^b(\mathbf{x,\hat{p}}) -\cos 2\phi' ~U^b(\mathbf{x,\hat{p}})  \right]\right\}\, .\label{eq:SSSSPV}
\end{align}
Notice that in Eqs.(\ref{eq:SPQ}), (\ref{eq:SPU}), and (\ref{eq:SPV}), the source terms proportional to $s_1$ and $s_2$ are linear in the perturbations. However, as we have just shown, such contributions at the end vanish, leaving therefore only the remaining source terms that are secondorder in the cosmological fluctuations.
Now, we can start to compare our results with previous calculations of this effect, i.e., Refs. \cite{Sawyer:2012gn, Montero-Camacho:2018vgs, Inomata:2018vbu}. For instance, in Ref. \cite{Sawyer:2012gn}, the time evolution of the CMB Stokes parameter $V$ obeys the following equation:
\beq
\frac{d}{d\eta} \Delta_ V^\gamma(\mathbf{k}) \propto  (1-\cos\theta)^2\left\{\Delta _Q^\gamma (\mathbf{k}) \left[\sin(2\phi) Q^b(\mathbf{p})-\cos(2\phi) U^b(\mathbf{p})\right]- \Delta _U^\gamma(\mathbf{k})\left[\sin(2\phi) U^b(\mathbf{p})+\cos(2\phi) Q^b(\mathbf{p})\right] \right\} \label{eq:V-dot Sawyer} \, ,
\eeq
where $\theta$ and $\phi$ are the polar angles between the observed and background photons. Comparing Eqs. \eqref{eq:SSSSPV} and \eqref{eq:V-dot Sawyer}, we find that our results are fully consistent with Ref. \cite{Sawyer:2012gn}, apart from different normalization conventions in the definition of CMB Stokes parameters.

We can show the consistency of our results also with Refs. \cite{Montero-Camacho:2018vgs, Inomata:2018vbu}. In these works, it is shown that the circular polarization of the radiation field is generally produced by Faraday conversion that occurs when a linearly polarized radiation propagates through a medium in which the axes perpendicular to the momentum of the incoming radiation have a different refraction index. 
In order to compare our results with Refs. \cite{Montero-Camacho:2018vgs, Inomata:2018vbu}, we need to expand the $Q$ and $U$ modes in Eq. \eqref{eq:density3} in terms of spin-weighted spherical harmonics $_{s}Y_{\ell m}$ as (see, e.g., Refs. \cite{Zaldarriaga:1996xe,Hu:1997hp})
\beq
(Q^b \pm iU^b)(\mathbf{x},\hat{\mathbf{p}})=P^ {b\pm} (\mathbf{x},\hat{\mathbf{p}})=\sum_{\ell=2}^{\infty}\sum_{m=-\ell}^\ell a^{(\pm 2)}_{\ell m}(\mathbf{x})~_{\pm 2}Y_{\ell m}(\hat{\mathbf{p}})=\sum_{\ell=2}^{\infty}\sum_{m=-\ell}^\ell  - (a^E_{\ell m}(\mathbf{x}) \pm  ia^B_{\ell m}(\mathbf{x}))~_{\pm 2}Y_{\ell m}(\hat{\mathbf{p}})\, ,\label{eq:qiu}
\eeq
where $a^E_{\ell m}$ and $a^B_{\ell m}$ denote the coefficients in the harmonic sphere expansion of the so-called $E$ and $B$ modes that give an alternative (rotationally invariant) description of the CMB linear polarization. 
Thus, we have
\beq
Q^b(\mathbf{x},\hat{\mathbf{p}})=\frac{1}{2}\sum_{\ell=2}^{\infty}\sum_{m=-\ell}^\ell [a_{\ell,m}^{(+2)}(\mathbf{x})~_{+2}Y_{\ell,m}(\hat{\mathbf{p}})+ a_{\ell,m}^{(-2)}(\mathbf{x})~_{-2}Y_{\ell,m}(\hat{\mathbf{p}})]\, ,
\eeq
and
\beq
U^b(\mathbf{x},\hat{\mathbf{p}})=\frac{1}{2i} \sum_{\ell=2}^{\infty}\sum_{m=-\ell}^\ell  [a_{\ell,m}^{(+2)}(\mathbf{x})~_{+2}Y_{\ell,m}(\hat{\mathbf{p}})- a_{\ell,m}^{(-2)}(\mathbf{x})~_{-2}Y_{\ell,m}(\hat{\mathbf{p}})]\, .
\eeq
Then, we also employ the fact that
\beq
a^E_{\ell,m}=  - \frac{1}{2}[a_{\ell,m}^{(+2)}+a_{\ell,m}^{(-2)}] \,, \qquad  \qquad a^B_{\ell,m}=  \frac{i}{2} [a_{\ell,m}^{(+2)}-a_{\ell,m}^{(-2)}] \, , 
\eeq
which follows from Eq. \eqref{eq:qiu}, together with the identities 
\beq
a^{E \ast}_{2,2}= a^{E}_{2,-2} \, , \qquad \qquad a^{B \ast}_{2,2}= a^{B}_{2,-2} \, ,
\eeq
that follow from the reality condition on the $E$ and $B$ modes.

Thus, after employing the angular decomposition of $Q$ and $U$ modes, we perform the momenta integration in Eqs. \eqref{eq:SSSSPQ}, \eqref{eq:SSSSPU}, and \eqref{eq:SSSSPV}, and we get\footnote{In deriving Eqs. \eqref{eq:PQ}, \eqref{eq:PU}, and \eqref{eq:PV}, we have used the following integral:
\begin{align}
\int_0^\infty \frac{\partial I_0(p)}{\partial p} \, p^4 ~dp=  I_0(p) p^4\Big|_0^{\infty} -  4 \int_0^\infty \, I_0(p) \, p^3 \, dp  = - 4 \int_0^\infty \frac{p^3 ~dp}{\textrm{exp}(p/T)-1}  =\frac{- 4 \pi^4 T^4}{15}=- 4 \pi^2 a_{rad}~ T^4 \, .
\end{align}
}
\begin{align}
\frac{d}{d\eta}{\Delta}_Q^\gamma(\mathbf{x},k_c)&= \frac{4}{5 \pi} \sqrt{\frac{\pi}{5}} \frac{ k^0_c \alpha^2}{  m_e^4} a_{rad} T^4_{CMB} \Delta_V^\gamma(\mathbf{x},k_c) \,   [ \textrm{Im} (a^E_{2,-2}(\mathbf{x}))- \textrm{Re}(a^B_{2,-2} (\mathbf{x}))]\, ,\label{eq:PQ} \\
\frac{d}{d\eta}{\Delta}_U^\gamma(\mathbf{x},k_c)&=- \frac{4}{5 \pi} \sqrt{\frac{\pi}{5}} \frac{ k^0_c \alpha^2}{  m_e^4} a_{rad} T^4_{CMB}  \Delta_V^\gamma(\mathbf{x},k_c) \,  [ \textrm{Re} (a^E_{2,-2}(\mathbf{x}))+\textrm{Im}(a^B_{2,-2}(\mathbf{x}))]\, ,\label{eq:PU}\\
\frac{d}{d\eta}{\Delta}_V^\gamma(\mathbf{x},k_c)&= \frac{4}{5 \pi} \sqrt{\frac{\pi}{5}} \frac{ k^0_c \alpha^2}{  m_e^4}  a_{rad} T^4_{CMB}\,  \left\{   \Delta_U^\gamma(\mathbf{x},k_c) [ \textrm{Re} (a^E_{2,-2}(\mathbf{x}))+\textrm{Im}(a^B_{2,-2}(\mathbf{x}))]\right. \nonumber\\
& \left. - \Delta_Q^\gamma(\mathbf{x},k_c)  [ \textrm{Im} (a^E_{2,-2}(\mathbf{x}))- \textrm{Re}(a^B_{2,-2}(\mathbf{x}))]\right\}\, ,\label{eq:PV}
\end{align}
where $a_{\textrm{rad}}=\pi^2/15$ is the radiation energy density constant and $T_{CMB}$ is the CMB temperature. Notice that our effect turns out to be proportional to CMB quadrupolar anisotropies of the linear polarization fields. This result is consistent with what we would have expected. In fact, it is well known that the linear polarization of a light beam is classically converted into circular polarization via Faraday conversion as it traverses a medium with different refraction indices along two orthogonal directions on the polarization plane of the propagating beam. Thus, since in our context the photon-photon forward scattering leads to an effective birefringent medium (induced by the target photons), then to realize Faraday conversion we need quadrupolar anisotropies in the distribution of the target (CMB) photons: in fact, only in this way can we get an effective medium with local different refraction indices along two orthogonal directions on the polarization plane of the incoming (CMB) photon. 

Notice also that Eq. \eqref{eq:PV} is consistent with Refs. \cite{Montero-Camacho:2018vgs, Inomata:2018vbu}, after ignoring the $a_{2,-2}^B$ term with respect to $a_{2,-2}^E$, which is equivalent to neglecting target (CMB) photons with $B$-mode polarization with respect to those with $E$-mode polarization. This assumption is well motivated by the fact that the amplitude of CMB $B$ modes is already constrained to be smaller than that of $E$ modes (see, e.g., Ref. \cite{Akrami:2018odb}), which thus will give the most relevant effects in our Faraday conversion. This is also equivalent to neglecting tensor perturbations from inflation as in the standard picture these represent the most important source of $B$-mode quadrupolar anisotropies.

We can obtain analogous equations for describing the Faraday conversion also in the general photon-photon forward-scattering case [\eqref{eq:general tensor}]. In fact, after using the spin-weighted spherical harmonic expansion and performing the momenta integrals in Eqs. \eqref{eq:SPQ}, \eqref{eq:SPU}, and \eqref{eq:SPV} with the same prescriptions as before, we find
\begin{align}
\frac{d}{d\eta}{\Delta}_Q^\gamma(\mathbf{x},k_c)&= (G_1+G_2+2G_4)  \frac{3}{\pi} \sqrt{\frac{\pi}{5}} k^0_c ~ a_{rad} T^4_{CMB} \Delta_V^\gamma(\mathbf{x},k_c) \,   [ \textrm{Im} (a^E_{2,-2}(\mathbf{x}))- \textrm{Re}(a^B_{2,-2}(\mathbf{x}))]\, ,\label{eq:G-SPQ} \\
\frac{d}{d\eta}{\Delta}_U^\gamma(\mathbf{x},k_c)&= -(G_1+G_2+2G_4)  \frac{3}{\pi} \sqrt{\frac{\pi}{5}}  k^0_c ~ a_{rad} T^4_{CMB}  \Delta_V^\gamma(\mathbf{x},k_c) \,  [ \textrm{Re} (a^E_{2,-2}(\mathbf{x}))+\textrm{Im}(a^B_{2,-2}(\mathbf{x}))]\, ,\label{eq:G-SPU} \\
\frac{d}{d\eta}{\Delta}_V^\gamma(\mathbf{x},k_c)&=  (G_1+G_2+2G_4) \frac{3}{\pi} \sqrt{\frac{\pi}{5}}  k^0_c ~  a_{rad} T^4_{CMB}\,  \left\{   \Delta_U^\gamma(\mathbf{x},k_c) [ \textrm{Re} (a^E_{2,-2}(\mathbf{x}))+\textrm{Im}(a^B_{2,-2}(\mathbf{x}))]\right. \nonumber\\
& \left. - \Delta_Q^\gamma(\mathbf{x},k_c)  [ \textrm{Im} (a^E_{2,-2}(\mathbf{x}))- \textrm{Re}(a^B_{2,-2}(\mathbf{x}))]\right\}\, .\label{eq:G-SPV}
\end{align}
Interestingly, the only difference between the general and the Euler-Heisenberg cases is in the overall coefficient $G_1+G_2+2G_4$, which in the Euler-Heisenberg case is fixed as in Eq.  \eqref{eq:EH_to_general}, but in the general case is undetermined.


\section{Power spectrum of circular polarization} \label{sec:4}

In this section, we derive the expression of the expected CMB circular polarization angular power spectrum induced by photon-photon forward scattering. We will assume Euler-Heisenberg interactions, but the final result will be generalized to any photon-photon interaction through Eq. \eqref{eq:EH_to_general}.
To this purpose, we first define the following quantities \cite{Zaldarriaga:1996xe}:
\begin{align}
\Delta_P^{\gamma \pm}=\Delta_ Q^\gamma \pm i \Delta_ U^\gamma \, ,
\end{align}
which encode CMB linear polarization and in Fourier space can be expressed in terms of rotationally invariant quantities, namely $E$ and $B$ modes.  Using Eqs. \eqref{eq:PQ} and \eqref{eq:PU} and assuming $a^B _{2,-2} \ll a^E _{2,-2} $, we get
\begin{align}
\frac{d}{d\eta}\Delta_P^{\gamma+}(\mathbf{x},k_c)&= \frac{4}{5 \pi}   \sqrt{\frac{\pi}{5}} \frac{k^0_c \alpha^2}{m_e^4}  a_{rad} T^4_{CMB} \, \Delta_V^\gamma(\mathbf{x},k_c)  \left[\textrm{Im} (a^E_{2,-2}(\mathbf{x})) + i \textrm{Re} (a^E_{2,-2}(\mathbf{x})) \right]\nonumber\\
&=i \frac{4}{5 \pi}   \sqrt{\frac{\pi}{5}} \frac{k^0_c \alpha^2}{m_e^4}  a_{rad} T^4_{CMB} \, \Delta_V^\gamma(\mathbf{x},k_c)  \, a^{E \ast }_{2,-2}(\mathbf{x}) \label{eq:BP}
\end{align}
and
\begin{align}
\frac{d}{d\eta}\Delta_P^{\gamma-}(\mathbf{x},k_c)= -i \frac{4}{5 \pi}  \sqrt{\frac{\pi}{5}} \frac{ k^0_c \alpha^2}{  m_e^4} a_{rad} T^4_{CMB} \Delta_V^\gamma(\mathbf{x},k_c) \, a^E_{2,-2}(\mathbf{x})\, .
\end{align}
These equations can be written in Fourier space as
\begin{align}
\frac{d}{d\eta}\Delta_P^{\gamma+}(\mathbf{K},k_c)&= i \frac{4}{5 \pi}  \sqrt{\frac{\pi}{5}} \frac{ k^0_c \alpha^2}{  m_e^4}  a_{rad} T^4_{CMB} \, \int \frac{d^3 \mathbf P}{(2 \pi)^3} \, \Delta_V^\gamma(\mathbf{K},k_c)  \, a^{E\ast }_{2,-2}(\mathbf{K} - \mathbf P) \, ,\label{eq:BP+} \\
\frac{d}{d\eta}\Delta_P^{\gamma-}(\mathbf{K},k_c)&= -i \frac{4}{5 \pi}  \sqrt{\frac{\pi}{5}} \frac{ k^0_c \alpha^2}{  m_e^4}  a_{rad} T^4_{CMB} \,  \int \frac{d^3 \mathbf P}{(2 \pi)^3} \,  \Delta_V^\gamma(\mathbf{K},k_c)  \, a^{E}_{2,-2}(\mathbf{K} - \mathbf P) \, ,\label{eq:BP-}
\end{align}
where $\mathbf{K}$ denotes the Fourier conjugate of $\mathbf{x}$.

Moreover, we also give the Fourier-space expression of the CMB $V$-mode polarization induced by Euler-Heisenberg interactions. From Eq. \eqref{eq:PV}, we get
\begin{align}
\frac{d}{d\eta}{\Delta}_V^\gamma(\mathbf{K},k_c)= \frac{4}{5 \pi} \sqrt{\frac{\pi}{5}} \frac{ k^0_c \alpha^2}{  m_e^4}  a_{rad} T^4_{CMB} \,   \int \frac{d^3 \mathbf P}{(2 \pi)^3}  \, \left[ \Delta_U^\gamma(\mathbf{K},k_c)  \textrm{Re} (a^E_{2,-2}(\mathbf{K} - \mathbf P))- \Delta_Q^\gamma(\mathbf{K},k_c)  \textrm{Im} (a^E_{2,-2}(\mathbf{K} - \mathbf P))\right]\, .\label{eq:BV}
\end{align}
Now, we need to implement in the equations of motion of the CMB polarization fields also the standard radiation transport terms, as given in the literature (see, e.g., Refs. \cite{Kosowsky:1994cy, Zaldarriaga:1996xe, Hu:1997hp}). These take into consideration the contributions of the photon-electron Thomson scattering and projection effects. Thus, the Boltzmann equations \eqref{eq:BP+}, \eqref{eq:BP-}, and \eqref{eq:BV} get modified into
\begin{align}
\frac{d}{d\eta} \Delta_P^{\gamma+} (\mathbf{K},k_c)+i K \mu \Delta_P^{\gamma+} (\mathbf{K},k_c)=&- \tau ^ \prime\left[- \Delta_P^{\gamma+} (\mathbf{K},k_c)+\frac{1}{2}\left(1-P_2(\mu)\right)\Pi(\mathbf{K}) + iA_L   \left[   \Delta_V^\gamma(\mathbf{K},k_c)  \,* \,  a^{E \ast }_{2,-2}(\mathbf{K}) \right]\right]\, ,\nonumber\\\label{eq:DBP+} \\
\frac{d}{d\eta} \Delta_P^{\gamma-} (\mathbf{K},k_c)+i K \mu \Delta_P^{\gamma-} (\mathbf{K},k_c)=& - \tau ^ \prime \left[- \Delta_P^{\gamma-} (\mathbf{K},k_c)+\frac{1}{2}\left(1-P_2(\mu)\right)\Pi(\mathbf{K}) - iA_L  \left[  \Delta_V^\gamma(\mathbf{K},k_c)  \,*\, a^{E}_{2,-2}(\mathbf{K}) \right] \right]\, ,\nonumber\\\label{eq:DBP-} \\
\frac{d}{d\eta} \Delta_V^\gamma (\mathbf{K},k_c)+i K \mu \Delta_V^\gamma (\mathbf{K},k_c)=& - \tau ^ \prime \Big\{ - \Delta_V^\gamma (\mathbf{K},k_c) + \frac{3}{2} \mu \Delta_{V1}^{\gamma} (\mathbf{K},k_c) + A_L  \left[   \Delta_U^\gamma(\mathbf{K},k_c) \,*\, \textrm{Re} (a^E_{2,-2}(\mathbf{K})) \right. \nonumber\\
& \left. - \Delta_Q^\gamma(\mathbf{K},k_c)  \,* \,\textrm{Im} (a^E_{2,-2}(\mathbf{K}))\right] \Big\} \, ,\label{eq:DBV}
\end{align}
where  a $*$ denotes convolution in Fourier space; $\tau'(\eta)$ is the so-called ``differential optical depth" of Thomson scattering, defined as
\begin{equation}\label{eq:optical}
\tau(\eta)= \int_\eta^{\eta_0} \, d\eta' \, a(\eta') n_e x_e \sigma_T\,,\,\,\,\,\,\,\,\,\quad \,\,\,\,\,\,\,\tau'(\eta)=  - a(\eta) n_e x_e \sigma_T\, ,
\end{equation}
with $n_e$ being the electron density, $x_e$ the ionization fraction, and $\sigma_T = (8 \pi/3) \, \alpha^2/m_e^2$ the Thomson cross section;
\begin{align}
\mu=\mathbf{\hat{K}}\cdot \mathbf{\hat{k}_c} \, 
\end{align}
is the cosine of the angle between the observed CMB photon and the Fourier mode $\mathbf{K}$; and 
\begin{align}
P_2(\mu) & = \frac{3 \mu^2 -1}{2}  \, , \\
A_L &= \frac{4}{5 \pi}  \sqrt{\frac{\pi}{5}} \frac{ k^0_c \alpha^2}{ m_e^4} \frac{a_{rad} (T_{CMB})^4}{a(\eta) n_e x_e \sigma_T}   = 10^{-2} \, \frac{2 \pi^2}{\zeta(3)}  \sqrt{\frac{\pi}{5}} \left(\frac{T_{CMB}^0}{m_e}\right) \left(\frac{k_{c}^0}{m_e}\right) \frac{(1+ z)^2}{x_e(z)} \left(\frac{n_\gamma}{n_e}\right)  \, , \label{eq:A_L}\\
\Pi &=\Delta_{I2}+\Delta_{Q0}+\Delta_{Q2}\, ,
\end{align}
where $\Delta_{I n}$, $\Delta_{Q n}$,  and $\Delta_{V n}$ represent the $n$th-order terms in the Legendre polynomial expansion of the corresponding quantities, $T^0_{CMB}$ denotes the CMB temperature today, $n_\gamma$ ($n_e$) is the photon (electron) number density, and $z$ is the redshift.

In analogy with the standard CMB radiation transport solutions, the differential equations \eqref{eq:DBP+}, \eqref{eq:DBP-}, and \eqref{eq:DBV} admit the following integral solutions:
\begin{align}
 \Delta_P^{\gamma+} (\eta_0,\mathbf{K}, k_c)=& \int_0^{\eta_0}  d\eta\,e^{iK\mu(\eta -\eta_0)-\tau} \tau ^ \prime(\eta) \left\{\frac{3}{4}(1-\mu^2)\, \Pi(K,\eta) + iA_L \,  \left[   \Delta_V^\gamma(\mathbf{K}, k_c) \,* \, a^{\ast E}_{2,-2}(\mathbf{K}) \right]\right\}\, , \label{eq:Delta+}\\
 \Delta_P^{\gamma-} (\eta_0,\mathbf{K}, k_c)=& \int_0^{\eta_0}  d\eta\,e^{iK\mu(\eta -\eta_0)-\tau} \tau ^ \prime(\eta) \left\{\frac{3}{4}(1-\mu^2)\, \Pi(K,\eta) - iA_L \,  \left[   \Delta_V^\gamma(\mathbf{K}, k_c) \, *\, a^{E}_{2,-2}(\mathbf{K}) \right]\right\}\, , \label{eq:Delta-} \\
 \Delta_V^\gamma (\eta_0,\mathbf{K}, k_c)=&\int_0^{\eta_0}  d\eta\,e^{iK\mu(\eta -\eta_0)-\tau} \tau ^ \prime(\eta) \biggl\{  \frac{3}{2} \mu \Delta_{V1}^{\gamma} (\mathbf{K},k_c) + A_L \, \Big[   \Delta_U^\gamma(\mathbf{K}, k_c) \,  *  \, \textrm{Re} (a^E_{2,-2}(\mathbf{K})) \nonumber \\
& - \Delta_Q^\gamma(\mathbf{K}, k_c) \, *  \, \textrm{Im} (a^E_{2,-2}(\mathbf{K}))\Big]\biggr\}\nonumber\\
=&\int_0^{\eta_0}  d\eta\,e^{iK\mu(\eta -\eta_0)-\tau} \tau ^ \prime(\eta) \left\{\frac{3}{2} \mu\Delta_{V1}^{\gamma} (\mathbf{K},k_c) + A_L \, \textrm{Im} \left[  a^{E\ast}_{2,-2}(\mathbf{K})\, * \, \Delta_p^{\gamma+}(\mathbf{K}, k_c)  \right]\right\}\, ,
\end{align}
where $\eta_0$ denotes the conformal time today with the condition $e^{-\tau(\eta_0)} = 1$. We have also assumed $e^{-\tau(0)} \approx 0$ as a first approximation. 

Now, following, e.g., Ref.~\cite{Zaldarriaga:1996xe}, in order to obtain the expected value of the $V$-mode polarization today in the $\hat n$ direction to the sky, we need to integrate over all the possible Fourier momenta as 
\begin{align} \label{eq:Vn}
 \Delta_V^\gamma (\hat{n}) = \int d^3\mathbf{K} \, \zeta(\mathbf{K})\, \tilde \Delta_V^\gamma (\eta_0,\mathbf{K}, \mu)\, ,
\end{align}
where $\zeta(\mathbf{K})$ is a random function used to describe the initial amplitude of primordial scalar perturbations from inflation.\footnote{We remind the reader that primordial tensor perturbations are neglected in our picture. In fact, they are observationally bound to have a much smaller amplitude, thus yielding a subdominant effect on the CMB polarization field.} After computing Eq. \eqref{eq:Vn}, we can define its harmonic sphere coefficients as
\begin{align}
 a^V_{\ell m}=\int d\Omega_n \,  Y^\ast_{\ell m} (\hat{{n}}) \Delta_V(\hat{{n}}) \, .\label{eq:a^V}
\end{align}
Then, the $V$-mode angular power spectrum reads
\begin{align}
 C^{VV}_\ell = \frac{1}{2\ell+1}\sum_{m} \braket{a^{V\ast}_{\ell m}\,\, a^V_{\ell m}}\, .
\end{align}
Therefore, inserting Eq. \eqref{eq:Vn} into \eqref{eq:a^V}, we get
\begin{align}
a^V_{\ell m}=  \int d\Omega_n \, Y^\ast_{\ell m}(\hat{n}) \int d^3\mathbf{K} \, \zeta(\mathbf{K}) \int_0^{\eta_0}  d\eta\, e^{iK\mu(\eta -\eta_0)-\tau} \tau ^ \prime(\eta) \left\{ \frac{3}{2} \mu  \Delta_{V1}^{\gamma} (\mathbf{K}) + A_L  \textrm{Im} \left[  a^{E\ast}_{2,-2}(\mathbf{K}) \,*\, \Delta_p^{\gamma+}(\mathbf{K},\mu)  \right]\right\}\label{eq:a^V2} \, ,
\end{align}
and thus the $V$-mode power spectrum reads
\begin{align}
C^{VV}_\ell =&\frac{1}{2\ell+1} \int d^3\mathbf{K}\,\mathcal{P}_\zeta (K) \nonumber\\
& \times \sum_m \left|\int d\Omega_n \, Y^\ast_{\ell m}(\hat{n}) \int_0^{\eta_0}  d\eta\, e^{iK\mu(\eta -\eta_0)-\tau}\tau ^ \prime(\eta) \left\{\frac{3}{2} \mu  \Delta_{V1}^{\gamma} (\mathbf{K}) + A_L  ( \textrm{Im} \left[ a^{E\ast}_{2,-2}(\mathbf{K}) \, * \,\Delta_p^{\gamma+}(\mathbf{K},\mu)  \right]\right\} \right|^2 \, ,  \label{eq:lv}
\end{align}
where $\mathcal{P}_{\zeta}(K)$, defined as
\beq
\braket{\zeta(\mathbf{K^\prime}) \zeta(\mathbf{K}))}=\delta^{(3)}(\mathbf{K^\prime}-\mathbf{K}) \mathcal{P}_{\zeta}(K) \, ,
\eeq
denotes the scalar primordial power spectrum from inflation. 

Equation \eqref{eq:lv} can be simplified by assuming that the circular polarization source terms are negligible in comparison with linear polarization terms.\footnote{This assumption is well motivated, as at present we do not have any observational evidence of circular polarization in the CMB, suggesting that the circular polarization signal, even if present, is much smaller than the linear polarization one. Moreover, we are here interested in a possible mechanism that, starting from initial vanishing $V$-mode polarization, does indeed produce it.} Therefore, Eq.  \eqref{eq:lv} reads
\begin{align}
 C^{VV}_\ell \simeq \frac{1}{2\ell+1}\int d^3\mathbf{K}\,\mathcal{P}_\zeta (K) \sum_m \left|\int d\Omega_n \, Y^\ast_{\ell m}(\hat{n})\,\int_0^{\eta_0} d\eta \,e^{iK\mu(\eta -\eta_0)} \, g(\eta) \, A_L \, \textrm{Im} \left[  a^{E\ast}_{2,-2}(\mathbf{K})\,* \, \Delta_P^{\gamma}(\mathbf{K},\mu)  \right] \right|^2 \label{eq:ClVV4}\, ,
\end{align}
where $g(\eta)=\tau ^ \prime e^{-\tau}$ is the so-called visibility function and
\begin{align}
 \Delta_P^{\gamma}(\mathbf{K},\mu)=& \int_0^{\eta}  d\eta'\,e^{iK\mu(\eta' -\eta)} g(\eta') \left[\frac{3}{4}(1-\mu^2)\, \Pi(K,\eta') \right]\, .\label{eq:Pg}
 \end{align}
Now, due to the product of the two visibility functions in Eqs. \eqref{eq:ClVV4} and \eqref{eq:Pg}, the latter takes the relevant contributions for $\eta' \simeq \eta$. Thus, Eq. \eqref{eq:ClVV4} becomes
\begin{align}
 C^{VV}_\ell \simeq &\frac{1}{2\ell+1}\int d^3\mathbf{K}\,\mathcal{P}_\zeta (K) \sum_m \left|\int d\Omega_n \, Y^\ast_{\ell m}(\hat{n})\,\int_0^{\eta_0} d\eta \,  g(\eta) \, \frac{3}{4} A_L (1-\mu^2)\, e^{i x \mu} \,\textrm{Im} \left[a^{E\ast}_{2,-2}(\mathbf{K})\,* \, \Pi(K,\eta)    \right] \right|^2 \nonumber \\
 =&\frac{1}{2\ell+1}\int d^3\mathbf{K}\,\mathcal{P}_\zeta (K) \sum_m \left|\int d\Omega_n \, Y^\ast_{\ell m}(\hat{n})\,\int_0^{\eta_0} d\eta \,  g(\eta) \, \frac{3}{4} A_L (1+ \partial_x^2)\, e^{i x \mu} \,\textrm{Im} \left[a^{E\ast}_{2,-2}(\mathbf{K})\,* \, \Pi(K)    \right] \right|^2 \label{eq:ClVV5}\, ,
\end{align}
where $x = K (\eta -\eta_0)$. Now, using the integral (see, e.g., Ref. \cite{Schmidt:2012ne})
\begin{align} \label{eq:int_Bessel}
\int d\Omega\, Y^\ast_{\ell m}(\theta, \phi)  \, e^{i x \mu} e^{i r \phi} (1 - \mu^2)^{|r|/2}= \sqrt{4 \pi (2 \ell + 1)}  \sqrt{\frac{(\ell+|r|)!}{(\ell - |r|)!}}\, i^r \, i^\ell \frac{j_\ell(x)}{x^{|r|}} \delta_{m r} \, ,
\end{align}
we can perform the angular integration in Eq. \eqref{eq:ClVV5}, obtaining
\begin{align}
 C^{VV}_\ell =& (4 \pi) \int d^3\mathbf{K} \, \mathcal{P}_\zeta (K) \left|\int_0^{\eta_0} d\eta \,  g(\eta) \, \frac{3}{4} A_L \Big[ j_\ell(x)+ j''_\ell(x) \Big]  \,\textrm{Im} \left[a^{E\ast}_{2,-2}(\mathbf{K})\,* \, \Pi(K)    \right] \right|^2  \nonumber\\
  =&(4 \pi) (\ell^4 - 2 \ell^3 + \ell^2 ) \int d^3\mathbf{K} \, \mathcal{P}_\zeta (K) \left|\int_0^{\eta_0} d\eta \,  g(\eta) \, \frac{3}{4} A_L \left[ \frac{ 2 \, j_{\ell+1}(x)}{\ell (\ell-1) x}+ \frac{j_\ell(x)}{x^2} \right]  \,\textrm{Im} \left[a^{E\ast}_{2,-2}(\mathbf{K})\,* \, \Pi(K)    \right] \right|^2  \label{eq:ClVV6}\, ,
\end{align}
where in the last step we have used  the differential equation satisfied by the spherical Bessel functions,
\begin{align} \label{eq:def_Bessel}
j''_\ell(x) + \left(\frac{2}{x}\right) j'_{\ell}(x) + \left(1- \frac{\ell(\ell+1)}{x^2} \right) j_\ell(x)  = 0 \, ,
\end{align}
together with the Bessel recurrence relation,
\begin{align} \label{eq:rec_Bessel}
j'_\ell(x) = - j_{\ell+1}(x) + \left(\frac{\ell}{x}\right) j_\ell(x) \, .
\end{align}
Finally, we rewrite the quantity
\beq \label{eq:Im}
\textrm{Im} \left[a^{E\ast}_{2,-2}(\mathbf{K})\,* \, \Pi(K,\eta)    \right] = \textrm{Im} \left[\int \frac{d^3 P}{(2 \pi)^3} \, a^{E\ast}_{2,-2}(\mathbf{K}-\mathbf{P}) \, \Pi(K)    \right] \, .
\eeq
Here, $a^{E}_{2,-2}(\mathbf{K}-\mathbf{P})$ can be expressed in terms of the same quantity in the frame where $\mathbf{K}-\mathbf{P}$ is aligned with the $z$ axis, as \cite{Montero-Camacho:2018vgs}
\begin{align}
a^{E}_{2,-2}(\mathbf{K}-\mathbf{P}) = D_{-2, 0}^{2}(\phi_{\mathbf{K}-\mathbf{P}}, \theta_{\mathbf{K}-\mathbf{P}}, 0) \, a^{E}_{2,0}(\mathbf{K}-\mathbf{P}\parallel z)  \, ,
\end{align}
where $D_{m, m'}^{\ell}(\alpha, \beta, \gamma)$ is the well-known Wigner rotation matrix \cite{wigner2012group}, and we have employed the fact that, since we have scalar perturbations, only the $m = 0$ term of $a^{E}_{2,m}(\mathbf{K}-\mathbf{P} \parallel z)$ gives a contribution. Thus, Eq. \eqref{eq:Im} reads
\begin{align}
\textrm{Im} \left[a^{E\ast}_{2,-2}(\mathbf{K})\,* \, \Pi(K,\eta)\right] =& - \frac{\sqrt 6}{4} \int  \frac{d^3P}{(2 \pi)^3} \,\sin^2\left(\theta_{\mathbf{K}-\mathbf{P}}\right)  \,  \sin\left(2\phi_{\mathbf{K}-\mathbf{P}} \right)  \, \, a^{E}_{2,0}(|\mathbf K-\mathbf{P}|) \,  \Pi(K) \,  ,
\end{align}
where we have used the fact that the quantity $a^{E}_{2,0}(\mathbf{K'}\parallel z)$ depends only on the wave number $K'$ due to the invariance of $E$ modes under rotations on the polarization plane. Thus, Eq. \eqref{eq:ClVV6} finally gives
\begin{align}
 C^{VV}_\ell =& (4 \pi) (\ell^4 - 2 \ell^3 + \ell^2 ) \int d^3\mathbf{K} \, \mathcal{P}_\zeta (K) \nonumber\\
 & \times \left|\int_0^{\eta_0} d\eta \,  g(\eta) \, \frac{3 \sqrt 6}{16} A_L \left[ \frac{ 2 \, j_{\ell+1}(x)}{\ell (\ell-1) x}+ \frac{j_\ell(x)}{x^2} \right]  \,  \int  \frac{d^3P}{(2 \pi)^3} \,\sin^2\left(\theta_{\mathbf{K}-\mathbf{P}}\right)  \,  \sin\left(2\phi_{\mathbf{K}-\mathbf{P}} \right)  \, \, a^{E}_{2,0}(|\mathbf K-\mathbf{P}|) \,  \Pi(K)   \right|^2  \label{eq:ClVV7}\, .
\end{align}
Now, in order to give an order-of-magnitude estimate on the amount of circular polarization produced by this effect, we employ in Eq. \eqref{eq:ClVV7} the expected level of CMB linear polarization. To this purpose, we define CMB $E$ and $B$ modes as \cite{Zaldarriaga:1996xe}
 \begin{align} 
\Delta_E^{\gamma}=&-\frac{1}{2}\left[\bar{\eth}^2 \Delta_P^{\gamma +}+\eth^2  \Delta_P^{\gamma -}\right] \, , \label{eq:E}\\
\Delta_B^{\gamma}=&\frac{i}{2}\left[\bar{\eth}^2 \Delta_P^{+}-\eth^2 \Delta_P^{-} \right]\, ,  \label{eq:B}
\end{align}
where $\eth$ and $\bar{\eth}$ are the so-called spin raising and lowering operators \cite{Zaldarriaga:1996xe}. Using these definitions, together with Eqs. \eqref{eq:Delta+} and \eqref{eq:Delta-}, we get \cite{Zaldarriaga:1996xe}
\begin{align}
\Delta_E^{\gamma}(\eta_0,\mathbf{K},\mu)=& - \int_0^{\eta_0}  d\eta\, g(\eta)\frac{3}{4}\, \Pi(K,\eta) \partial_\mu^2 \left[ (1-\mu^2)^2 \, e^{iK\mu(\eta -\eta_0)}\right] \, ,   \nonumber\\
=&  \int_0^{\eta_0}  d\eta\, g(\eta)\frac{3}{4}\, \Pi(K,\eta) \, Q(x) \left( x^2 \, e^{i x \mu} \right) \, , \\
\Delta_B^{\gamma}(\eta_0,\mathbf{K},\mu) =& \, 0 \, ,
\end{align}
where again $x= K(\eta -\eta_0)$ and $Q(x) = (1+\partial_x^2)^2$. Here, we have neglected the backreaction terms due to the coupling with circular polarization. Now, following again Ref. \cite{Zaldarriaga:1996xe}, the expected $E$-mode angular power spectrum is given by
\begin{align}
 C^{EE}_\ell =\frac{1}{2\ell+1}\sum_{m} \braket{a^{E\ast}_{\ell m}\,\, a^E_{\ell m}}\, ,
\end{align}
where
\begin{align}
 a^E_{\ell m}= \left[\frac{(\ell -2)!}{(\ell + 2)!}\right]^{\frac{1}{2}} \int d\Omega_n \,  Y^\ast_{\ell m} (\hat{n}) \Delta_E(\hat{n}) \, ,
\end{align}
and
\begin{align}
 \Delta_E^\gamma (\hat{n}) = \int d^3\mathbf{K} \, \zeta(\mathbf{K})\, \tilde \Delta_E^\gamma (\eta_0,\mathbf{K},\mu)\, .
\end{align}
 Thus, we get
\begin{align}
 C^{EE}_\ell =&\frac{1}{2\ell+1}  \left[\frac{(\ell -2)!}{(\ell + 2)!}\right] \, \int d^3\mathbf{K}\,\mathcal{P}_\zeta (K) \sum_m \left|\int d\Omega_n \, Y^\ast_{\ell m}(\hat{n})\,\int_0^{\eta_0} d\eta\, g(\eta)\frac{3}{4}\, \Pi(K,\eta)  \, Q(x) \left( x^2 \, e^{i x \mu} \right) \right|^2  \nonumber \\
 =&(4 \pi) \left[\frac{(\ell -2)!}{(\ell + 2)!}\right]\int d^3 \mathbf{K}\,\mathcal{P}_\zeta (K) \left|\int_0^{\eta_0} d\eta\, g(\eta)\frac{3}{4}\, \Pi(K,\eta)  \, Q(x) \, \left(x^2 \, j_\ell(x) \right) \right|^2 \nonumber \\
 =&(4 \pi) \left(\ell^4 + 2 \ell^3 - \ell^2 - 2 \ell \right) \int d^3\mathbf{K}\,\mathcal{P}_\zeta (K) \left|\int_0^{\eta_0} d\eta\, g(\eta)\frac{3}{4}\, \Pi(K,\eta)  \,  \frac{j_\ell(x)}{x^2} \right|^2 \, , \label{eq:CEE}
\end{align}
where we have used again Eqs. \eqref{eq:int_Bessel}, \eqref{eq:def_Bessel}, and \eqref{eq:rec_Bessel}.

Now, by the matching between Eqs. \eqref{eq:ClVV7} and \eqref{eq:CEE}, we get the following approximate relation between the circular and linear CMB polarization fields:
\begin{align} \label{eq:V_toE}
 C^{VV}_\ell \approx  \left[\bar A^2_L(\eta_{rec}) \, C_2^{EE}(\eta_{rec}) \right] \, C^{EE}_\ell \, ,
\end{align} 
which holds apart from $\sim\mathcal O(1)$ coefficients. Here,  $\bar A_L(\eta_{rec})$ denotes the redshift-averaged value of $A_L(\eta)$ for an average CMB comoving frequency mode $\bar k^0_c = \pi^4/(30\zeta(3)) \,T^0_{CMB}$. This is estimated by evaluating Eq. \eqref{eq:A_L}  for $k^0_c = \bar k^0_c$ as
\begin{align}
A_L(\bar k^0_c) = 10^{-3} \, \frac{2 \pi^6}{3 \zeta^2(3)}  \sqrt{\frac{\pi}{5}} \left(\frac{T_{CMB}^0}{m_e}\right)^2 \frac{(1+ z)^2}{x_e(z)} \left(\frac{n_\gamma}{n_e}\right)  \, , 
\end{align}
and taking the following values for the constant parameters:
\begin{align}
\frac{n_\gamma}{n_e} =  2 \times 10^{9} \, , \qquad \qquad m_e = 5 \times 10^5 \, \mbox{ eV} \, , \qquad \qquad T_{CMB}^0 = 3.1 \times 10^{-4} \, \mbox{ eV} \, ,
\end{align}
and the following redshift average:
\begin{align}
 \frac{1}{z_{rec}}\, \int_{0}^{z_{rec}} d z \,  \frac{(1+ z)^2}{x_e(z)} \simeq 9 \times 10^7 \, ,
 \end{align}
 where $z_{rec}$ indicates the redshift at the recombination epoch. Therefore, we obtain
\begin{align}
\bar A_L(\eta_{rec}) \simeq   2.4\times 10^{-2} \,  .
\end{align}
Also, we know that the relative amplitude of the CMB $E$-mode polarization quadrupole left imprinted by scalar perturbations is of the order \cite{Aghanim:2019ame}
\begin{align}
\sqrt{C_2^{EE}} = \frac{\Delta_{E_2}}{T_0} \sim 10^{-6} \, .
\end{align}
Thus, Eq. \eqref{eq:V_toE} reads
\begin{align} \label{eq:V_toE2}
 C^{VV}_\ell \approx 10^{-16} \, C^{EE}_\ell \, .  
\end{align} 
This result suggests that the number of $V$ modes produced by this effect is much smaller than the level of linear polarization that we actually observe in the CMB, in such a way that our procedure of neglecting the backreaction of $V$-mode source terms on linear polarization is a consistent and very good approximation. Moreover, the rms value of $V$ modes is given approximately by 
\begin{equation} \label{eq:V_toE2_amplitudes}
V_{\rm rms} \simeq \left(\frac{1}{2 \pi^2} \int d^2\ell \, C_\ell^{VV} \right)^{1/2} \approx \left(10^{-16} \, \frac{1}{2 \pi^2}\int d^2\ell \, C_\ell^{EE} \right)^{1/2} = 10^{-8} E_{\rm rms} \sim 10^{-14} \,{ \rm K}  \, .
\end{equation} 
Notice that by employing Eq. \eqref{eq:EH_to_general}, we can express Eq. \eqref{eq:V_toE2_amplitudes} in terms of the $G_i$ general coefficients as
\begin{align}  \label{eq:V_toE2_amplitudes_general}
V_{\rm rms} \approx  10^{-8} \left(G_1 + G_2 + 2 G_4 \right) \frac{15 m_e^4}{4 \alpha^2} \, E_{\rm rms} \, ,
\end{align} 
which gives the order of magnitude of the expected amplitude of $V$ modes from the photon-photon forward scattering mediated by a generic interaction. 

Now, confronting our result in Eq. \eqref{eq:V_toE2_amplitudes} with the one found in Ref. \cite{Sadegh:2017rnr}, it turns out that our result is about 4 orders of magnitude smaller. This discrepancy may be explained by the following considerations: First, in Ref. \cite{Sadegh:2017rnr}, the coupling between the observed CMB linear and circular polarization is realized through the CMB background intensity field. But, as is also emphasized in Ref. \cite{Montero-Camacho:2018vgs}, this is not possible, because in such a case the $f_2^0$ parameter of Eq. (49) in Ref. \cite{Sadegh:2017rnr} would identically vanish once we perform the underlying angular integrals. As we have explicitly shown in our work, Faraday conversion is possible only if the coupling is realized through the linear polarization field of the background photons. This leads to a 6-order-of-magnitude discrepancy. Moreover, we notice that in Ref. \cite{Sadegh:2017rnr}, the matching between the linear and circular polarization fields is made by taking the average value of the parameter $\eta_{EH}(z)$ in their Eq. (49), which is related to our $A_L(\eta)$ apart from constant coefficients.  These constant coefficients are such that $\eta^{a v}_{EH}$ is 2 orders of magnitude smaller than $\bar A_L(\eta_{rec})$.

Thus, the 4-order-of-magnitude difference is explained by the exchange $Q(\mathbf p) (U(\mathbf p))\longleftrightarrow I(\mathbf p)$ between our Eq. \eqref{eq:SSSSPV} and Eq. (12) in  Ref. \cite{Sadegh:2017rnr}, together with the exchange $\bar A_L(\eta_{rec}) \longleftrightarrow \eta^{a v}_{EH}$.

\section{Polarization mixing from photon and massive spin-1 particle forward scattering} \label{sec:5}

In this section, we investigate a new viable way to get circular polarization out of the forward scattering between CMB photons and massive spin-1 particles. An example of such cosmological candidates are the so-called \textit{hidden photons}. These massive bosons are present in extensions of the Standard Model of particle physics containing a general new hidden U(1) gauge group \cite{Nelson:2011sf, Arza:2017phd, Daido:2018dmu, Foldenauer:2019dai}. The dominant interaction between the conventional photons and the hidden photons is realized through the gauge kinetic mixing between them. In the literature, different methods have been proposed to constrain the coupling and mass of hidden photons using astrophysical and cosmological observations \cite{An:2014twa, Abe:2018owy, Knirck:2018ojz, Baryakhtar:2018doz, Bauer:2018onh, Danilov:2018bks, Kovetz:2018zes, Brun:2019kak, Randall:2019zol, Hambye:2019dwd, Nguyen:2019xuh,Phipps:2019cqy, Raaijmakers:2019hqj, Chaudhuri:2019ntz, Kopylov:2019xqw, Demidov:2018odn}. Here, we will consider only a generic photon-spin-1-massive-particle scattering, leaving the extension of our final set of equations for specific cases to future work.

First of all, it is well known that a massive spin-1 field satisfies the so-called Proca equation \cite{Schwartz:2013pla}, with a mass term that explicitly breaks gauge invariance. The polarization vector of such a field involves three independent components. Moreover, the polarization field is characterized by eight parameters which describe all the possible independent polarization states (see Appendix \ref{appendix_C} for a brief review). In particular, the polarization matrix of a massive photon can be written as \cite{Lakin-1955, Gomes-1981, Ramachandran-1980}
\beq 
\rho_{ij}=\frac{tr(\rho)}{3}\left[\mathbb{1}_3 \, +\sum_{i=1}^{8} \lambda_i T_i  \right]\, ,
\eeq
where $\rho_{ij}$ is a $3\times 3$ matrix, $\lambda_i$ are the generators of the SU(3) group, $tr(\rho) = I$ is the intensity field describing unpolarized massive photons, and $T_i$ are eight generalized Stokes parameters describing the polarization state of a given system. However, since there is little knowledge about the physical polarization states of massive spin-1 particles, here we will focus only on their intensity, assuming an unpolarized background.

As in the previous examples, we begin by writing down the scattering amplitude. The general scattering amplitude of two generic spin-1 particles has been found by Ref. \cite{Costantini:1971} in the following form:
\begin{align} \label{eq:amplitudemassivespin-1}
M_{\mu \nu \rho \sigma}(1234)=&A_1(1234) I^{(1)}_{\mu \nu \rho \sigma} (1234)+A_2(1234) I^{(2)}_{\mu \nu \rho \sigma} (1234)+A_3(1234) I^{(3)}_{\mu \nu \rho \sigma} (1234)\nonumber\\
&+A_4(1234) I^{(4)}_{\mu \nu \rho \sigma} (1234)+A_5(1234) I^{(5)}_{\mu \nu \rho \sigma} (1234)\, ,
\end{align}
where $A_i$ are Lorentz-invariant coefficients and $ I^{(i)}_{\mu \nu \rho \sigma} $ are five independent gauge-invariant tensors that must be determined. After applying crossing and gauge-invariant symmetries, the $I^{(i)}_{\mu \nu \rho \sigma}$ tensors are given by \cite{Costantini:1971}
\begin{align}
 I^{(1)}_{\mu \nu \rho \sigma} (1234)=&\frac{1}{8}\left\{ p^1_\nu p^2_\mu p^3_\sigma p^4_\lambda -\delta_{\lambda \sigma} p^1_\nu p^2_\mu (p^3 \cdot p^4) - \delta_{\mu \nu} p^3_\sigma p^4_\lambda (p^1 \cdot p^2) +\delta_{\mu \nu} \delta_{\lambda \sigma} (p^1 \cdot p^2)  (p^3 \cdot p^4) \right\}\, ,\\
 I^{(2)}_{\mu \nu \rho \sigma} (1234) =&\frac{1}{8}\left\{p^1_\sigma p^2_\mu p^3_\nu p^4_\lambda+p^1_\nu p^2_\lambda p^3_\sigma p^4_\mu+\delta_{\nu \lambda} p^2_\mu p^3_\sigma (p^1 \cdot p^4)-\delta_{\nu \lambda} p^1_\sigma p^2_\mu (p^3 \cdot p^4)-\delta_{\lambda \sigma} p^2_\mu p^3_\nu (p^1 \cdot p^4)  \right. \nonumber\\
&\left.  - \delta_{\mu \sigma} p^1_\nu p^2_\lambda (p^3 \cdot p^4) -\delta_{\lambda \sigma} p^1_\nu p^4_\mu (p^2 \cdot p^3)  +\delta_{\lambda \sigma} p^3_\nu p^4_\mu (p^1 \cdot p^2) -\delta_{\nu \lambda} p^3_\sigma p^4_\mu (p^1 \cdot p^2) \right. \nonumber\\
& \left. -\delta_{\mu \sigma} p^3_\nu p^4_\lambda (p^1\cdot p^2)-\delta_{\mu  \nu} p^2_\lambda p^3_\sigma (p^1 \cdot p^4)-\delta_{\mu  \nu} p^1_\sigma p^4_\lambda (p^2 \cdot p^3)+\delta_{\mu  \nu} p^1_\sigma p^2_\lambda (p^3 \cdot p^4) \right. \nonumber\\
& \left. +\delta_{\mu \sigma} p^1_\nu p^4_\lambda (p^2 \cdot p^3)+\delta_{\mu \nu} \delta_{\lambda \sigma} (p^1 \cdot p^4)  (p^2 \cdot p^3)+\delta_{\mu \sigma} \delta_{\nu \lambda} (p^1 \cdot p^2)  (p^3 \cdot p^4) \right\}\, ,\\
 I^{(3)}_{\mu \nu \rho \sigma} (1234)=& -\frac{1}{2p^3 p^4 }\left\{p^1_\nu p^2_\mu p^1_\lambda p^3_\sigma (p^1 \cdot p^4) +p^1_\nu p^2_\mu p^1_\sigma p^4_\lambda (p^1 \cdot p^3) -p^1_\nu p^2_\mu p^1_\lambda p^1_\sigma (p^3 \cdot p^4)-\delta_{\lambda \sigma} p^1_\nu p^2_\mu  (p^1 \cdot p^3) (p^1 \cdot p^4) \right. \nonumber\\
& \left.-\delta_{\mu \nu} p^1_\lambda p^3_\sigma  (p^1 \cdot p^2) (p^1 \cdot p^4)+\delta_{\mu \nu} p^1_\lambda p^1_\sigma  (p^1 \cdot p^2) (p^3 \cdot p^4) -\delta_{\mu \nu} p^1_\sigma p^4_\lambda  (p^1 \cdot p^2) (p^1 \cdot p^3)\right. \nonumber \\
& \left.+\delta_{\mu \nu}\delta_{\lambda \sigma}  (p^1 \cdot p^2)  (p^1 \cdot p^3) (p^1 \cdot p^4)\right\}\, , \\
 I^{(4)}_{\mu \nu \rho \sigma} (1234)=& -\frac{1}{2p^3 p^4 }\left\{p^1_\nu p^2_\mu p^2_\lambda p^3_\sigma (p^1 \cdot p^4)+p^1_\nu p^2_\mu p^1_\sigma p^4_\lambda (p^2 \cdot p^3)-p^1_\nu p^2_\mu p^2_\lambda p^1_\sigma (p^3 \cdot p^4)-\delta_{\lambda \sigma} p^1_\nu p^2_\mu  (p^2 \cdot p^3) (p^1 \cdot p^4) \right. \nonumber\\
& \left. -\delta_{\mu \nu} p^2_\lambda p^3_\sigma  (p^1 \cdot p^2) (p^1 \cdot p^4)+\delta_{\mu \nu} p^2_\lambda p^1_\sigma  (p^1 \cdot p^2) (p^3 \cdot p^4)-\delta_{\mu \nu} p^1_\sigma p^4_\lambda  (p^1 \cdot p^2) (p^2 \cdot p^3) \right. \nonumber\\
& \left.+\delta_{\mu \nu}\delta_{\lambda \sigma}  (p^1 \cdot p^2)  (p^2 \cdot p^3) (p^1\cdot p^4)\right\}\, ,\\
I^{(5)}_{\mu \nu \rho \sigma} (1234)=& \frac{1}{3 p^2 p^4 }\left\{p^1_\sigma p^2_\mu p^3_\nu p^1_\lambda (p^2 \cdot p^4)+p^2_\sigma p^3_\mu p^2_\lambda p^1_\nu (p^1 \cdot p^4)-p^1_\sigma p^3_\mu p^2_\lambda p^1_\nu (p^2 \cdot p^4)-p^2_\sigma p^2_\mu p^3_\nu p^1_\lambda (p^1 \cdot p^4) \right. \nonumber\\
& \left. +\delta_{\nu \lambda} p^2_\sigma p^2_\mu  (p^1 \cdot p^4) (p^1 \cdot p^3)+\delta_{\mu \lambda} p^2_\sigma p^3_\nu  (p^1 \cdot p^4) (p^1 \cdot p^2) - \delta_{\mu \lambda} p^2_\sigma p^1_\nu  (p^1 \cdot p^4) (p^2 \cdot p^3)\right. \nonumber\\
& \left.+\delta_{\mu \nu} p^2_\sigma p^1_\lambda  (p^1 \cdot p^4) (p^2 \cdot p^3)-\delta_{\lambda \nu} p^2_\sigma p^3_\mu  (p^1 \cdot p^4) (p^1 \cdot p^2)+\delta_{\mu \nu} p^1_\sigma p^2_\lambda  (p^1 \cdot p^3) (p^2 \cdot p^4)\right. \nonumber\\
& \left. -\delta_{\mu \nu} p^1_\sigma p^1_\lambda  (p^2 \cdot p^3) (p^2 \cdot p^4)-\delta_{\nu \lambda} p^1_\sigma p^2_\mu  (p^1 \cdot p^3) (p^2 \cdot p^4)-\delta_{\mu \lambda} p^1_\sigma p^3_\nu  (p^1 \cdot p^2) (p^2 \cdot p^4) \right. \nonumber\\
& \left.+\delta_{\mu \lambda} p^1_\sigma p^1_\nu  (p^2 \cdot p^3) (p^2 \cdot p^4)-\delta_{\mu \nu} p^2_\sigma p^2_\lambda  (p^1 \cdot p^4) (p^1 \cdot p^3)+\delta_{\lambda \nu} p^1_\sigma p^3_\mu  (p^1 \cdot p^2) (p^2 \cdot p^4)\right\}\, .
\end{align}
Now, inserting the invariant amplitude [Eq. \eqref{eq:amplitudemassivespin-1}] into the quantum Boltzmann equation, and imposing the massless condition for CMB photons and the massive condition for massive spin-1 fields, we find the following expressions for the time evolution of the CMB Stokes parameters:
\begin{align}
\frac{d}{dt}{\Delta}_I^\gamma(\mathbf{x,k}_c)=& \, 0 \, , \\
\frac{d}{dt}{\Delta}_Q^\gamma(\mathbf{x,k}_c) =& \frac{1}{6 k^2 p}\Delta_V^\gamma(\mathbf{x,k}_c)  \int \frac{d^3\mathbf{p}}{(2\pi)^3 2p^0}\, \left[\frac{p^0}{8}\frac{\partial I_0(p)}{\partial p^0}\right] \,\left [L_0\, I^b(\mathbf{\hat{p}}) \label{eq:massiveQ}\right] \, ,\\
\frac{d}{dt}{\Delta}_U^\gamma(\mathbf{x,k}_c) =& -\frac{1}{12 k^2 p} \Delta_V^\gamma(\mathbf{x,k}_c)  \int \frac{d^3\mathbf{p}}{(2\pi)^3 2p^0} \, \left[\frac{p^0}{8}\frac{\partial I_0(p)}{\partial p^0}\right]\,\left[M_0\, I^b(\mathbf{\hat{p}}) \label{eq:massiveU}\right] \, ,\\
\frac{d}{dt}{\Delta}_V^\gamma(\mathbf{x,k}_c) =&\frac{1}{12 k^2 p} \int \frac{d^3\mathbf{p}}{(2\pi)^3 2p^0} \, \left[\frac{p^0}{8}\frac{\partial I_0(p)}{\partial p^0}\right]\,I^b(\mathbf{\hat{p}}) \left[-2L_0\,\Delta_Q^\gamma(\mathbf{x,k}_c) + M_0 \, \Delta_U^\gamma(\mathbf{x,k}_c) \right]\label{eq:massiveV} \, ,
\end{align}
where $I^b(\mathbf{p})$ labels the intensity of the background massive spin-1 field. The parameters $L_0$ and $M_0$ are given in Appendix \ref{appendix_D}. It is straightforward to check that the coefficients $L_0$ and $M_0$ vanish in a general reference frame. This result shows that just like the photon-photon scattering case, the CMB $V$-mode polarization does not couple with the intensity of massive spin-1 photons, but only with their polarization fields. 


\section{Conclusion} \label{sec:6}

In this paper we have studied the generation of CMB $V$ modes from photon-photon forward scattering using a quantum Boltzmann equation formalism. We have derived a set of general Eqs. \eqref{eq:SPQ2}, \eqref{eq:SPU2}, and \eqref{eq:SPV2} describing the \textit{conversion} of CMB linear polarization into circular polarization. Then, we specialized to the case of Euler-Heisenberg interactions and derived the consequent amplitude of $V$ modes produced. Our final estimation, in Eq. \eqref{eq:V_toE2_amplitudes}, is in line with previous literature \cite{Montero-Camacho:2018vgs,Inomata:2018vbu} and corrects the computations made by a previous paper \cite{Sadegh:2017rnr}, which used the same formalism adopted here.

Since the amplitude of CMB circular polarization is expected to be almost 8 orders of magnitude smaller than the amplitude of CMB linear polarization, we have no hope to observe soon such a signature with CMB experiments. 

Moreover, throughout this paper we have provided very general expressions extending the computations described above and the corresponding expected number of $V$ modes in the case of a generic photon-photon interaction, not relying on any specific fundamental interaction [Eq. \eqref{eq:V_toE2_amplitudes_general}]. 
Finally, we have investigated the possibility to get some $V$ modes from the forward scattering between CMB photons and spin-1 massive particles. Our final set of equations \eqref{eq:massiveQ}, \eqref{eq:massiveU}, and \eqref{eq:massiveV} confirms that only polarized spin-1 massive particles can couple linear to circular polarization in the CMB. These latter results provide the basis to further investigate new cosmological signatures of physics beyond the standard model of particle physics. We leave further study in this direction for future research.

\section*{Acknowledgments}

Part of the computations have been done using the FeynCalc \textit{Mathematica} package \cite{Mertig:1990an, Shtabovenko:2016sxi}. M. Z. acknowledges financial support by the University of Padova under the MSCA Seal of Excellence @UniPD program. N. B., S. M ., and G. O. acknowledge partial financial support by ASI Grant No. 2016-24-H.0.

\begin{appendices}

\section{coefficients of Stokes parameters in the case of general photon-photon interaction} \label{appendix_A}

Here, we provide the explicit expression of the coefficients in Eqs. \eqref{eq:SPQ}, \eqref{eq:SPU}, and \eqref{eq:SPV}:
\begin{align}
g_1=&(G_1+G_2+2G_4)\left\{(k \cdot p)^2\left[(\epsilon_2(p)\cdot \epsilon_1(k))(\epsilon_2(p) \cdot \epsilon_2(k))+(\epsilon_1(p) \cdot \epsilon_1(k))(\epsilon_1(p) \cdot \epsilon_2(k))\right]\right.\nonumber\\
&\left.-(k\cdot p)(k \cdot \epsilon_2(p))\left[(p \cdot \epsilon_2(k))(\epsilon_2(p) \cdot \epsilon_1(k))+(p \cdot \epsilon_1(k))(\epsilon_2(p) \cdot \epsilon_2(k))\right]\right.\nonumber\\
&\left.- (k\cdot p)(k \cdot \epsilon_1(p))\left[(p \cdot \epsilon_2(k))(\epsilon_1(p) \cdot \epsilon_1(k))+(p \cdot \epsilon_1(k))(\epsilon_1(p) \cdot \epsilon_2(k))\right]\right.\nonumber\\
& \left. + (p \cdot \epsilon_1(k))(p \cdot \epsilon_2(k))\left[(k \cdot \epsilon_1(p))^2+(k \cdot \epsilon_2(p))^2\right]\right\}\, , \\
g_2=&(G_1+G_2+2G_4)\left\{ (k \cdot p)^2\left[(\epsilon_1(p) \cdot \epsilon_1(k))(\epsilon_2(p) \cdot \epsilon_2(k)) +(\epsilon_2(p) \cdot \epsilon_1(k))(\epsilon_1(p) \cdot \epsilon_2(k))\right]\right.\nonumber\\
&\left.-(k \cdot p)(k \cdot \epsilon_2(p))\left[(p \cdot \epsilon_2(k))(\epsilon_1(p) \cdot \epsilon_1(k))+(p \cdot \epsilon_1(k))(\epsilon_1(p) \cdot \epsilon_2(k))\right]\right.\nonumber\\
&\left. -(k \cdot p)(k \cdot \epsilon_1(p))\left[(p \cdot\epsilon_2(k))(\epsilon_2(p) \cdot \epsilon_1(k))+(p \cdot \epsilon_1(k))(\epsilon_2(p) \cdot \epsilon_2(k))\right]\right.\nonumber\\
&\left.+2(k \cdot \epsilon_1(p))(k \cdot \epsilon_2(p))(p \cdot \epsilon_1(k))(p \cdot \epsilon_2(k))\right\}\, , \\
g_3=&(G_1+G_2+2G_4)\left\{(k \cdot p)^2\left[(\epsilon_1(p) \cdot \epsilon_1(k))(\epsilon_1(p) \cdot \epsilon_2(k)) - (\epsilon_2(p) \cdot \epsilon_1(k))(\epsilon_2(p) \cdot \epsilon_2(k))\right]\right.\nonumber\\
&\left.+(k \cdot p)(k \cdot \epsilon_2(p))\left[(p \cdot \epsilon_2(k))(\epsilon_2(p) \cdot \epsilon_1(k))+(p \cdot \epsilon_1(k))(\epsilon_2(p) \cdot\epsilon_2(k))\right]\right.\nonumber\\
&\left.- (k \cdot p)(k \cdot \epsilon_1(p))\left[(p \cdot \epsilon_2(k))(\epsilon_1(p) \cdot \epsilon_1(k))+(p \cdot \epsilon_1(k))(\epsilon_1(p) \cdot\epsilon_2(k))\right]\right.\nonumber\\
& \left. + (p \cdot \epsilon_1(k))(p \cdot \epsilon_2(k))\left[(k \cdot \epsilon_1(p))^2 - (k \cdot \epsilon_2(p))^2\right]\right\}\, , \\
g_4=&(G_1-G_2)\left\{(k \cdot p)^2\left[(\epsilon_1(p) \cdot \epsilon_1(k))(\epsilon_2(p) \cdot \epsilon_2(k))-(\epsilon_1(p) \cdot \epsilon_2(k))(\epsilon_2(p) \cdot \epsilon_1(k))\right]\right.\nonumber\\
&\left.-(k \cdot p)(k \cdot \epsilon_2(p))\left[(p \cdot \epsilon_2(k))(\epsilon_1(p) \cdot \epsilon_1(k))- (p \cdot \epsilon_1(k))(\epsilon_1(p) \cdot \epsilon_2(k))\right]\right.\nonumber\\
&\left.+ (k \cdot p)(k \cdot \epsilon_1(p))\left[(p \cdot \epsilon_2(k))(\epsilon_2(p) \cdot \epsilon_1(k))- (p \cdot \epsilon_1(k))(\epsilon_2(p) \cdot \epsilon_2(k))\right]\right\}\, , \\
g_5=&(G_1+G_2+2G_4)\left\{(k \cdot p)^2\left[ (\epsilon_1(p) \cdot \epsilon_1(k))^2-(\epsilon_1(p) \cdot \epsilon_2(k))^2 +(\epsilon_2(p) \cdot \epsilon_1(k))^2-(\epsilon_2(p) \cdot \epsilon_2(k))^2\right] \right. \nonumber\\
& \left. +2(k \cdot p)(k \cdot \epsilon_1(p))\left[(p \cdot \epsilon_2(k))(\epsilon_1(p) \cdot \epsilon_2(k))-(p \cdot \epsilon_1(k))(\epsilon_1(p) \cdot \epsilon_1(k))\right] \right. \nonumber\\
&\left.+2 (k \cdot p)(k \cdot \epsilon_2(p)) \left[(p \cdot \epsilon_2(k))(\epsilon_2(p) \cdot \epsilon_2(k))-(p \cdot \epsilon_1(k))(\epsilon_2(p) \cdot \epsilon_1(k))\right]\right. \nonumber\\
& \left. + \left((k \cdot \epsilon_1(p))^2+(k \cdot \epsilon_2(p))^2\right)\left[(p \cdot \epsilon_1(k))^2-(p \cdot \epsilon_2(k))^2\right]\right\}\, , \\
g_6=& 2(G_1+G_2+2G_4)\left\{ (k \cdot p)^2 \left[  (\epsilon_2(p) \cdot \epsilon_1(k))(\epsilon_1(p) \cdot \epsilon_1(k)) - (\epsilon_1(p) \cdot \epsilon_2(k))(\epsilon_2(p) \cdot \epsilon_2(k)) \right]\right. \nonumber\\
& \left. +(k \cdot p)(k \cdot \epsilon_1(p))\left[(p \cdot \epsilon_2(k))(\epsilon_2(p) \cdot \epsilon_2(k)) - (p \cdot \epsilon_1(k))(\epsilon_2(p) \cdot \epsilon_1(k))\right] \right. \nonumber \\
& \left. +(k \cdot p)(k \cdot \epsilon_2(p))\left[ (p \cdot \epsilon_2(k))(\epsilon_1(p) \cdot \epsilon_2(k)) - (p \cdot \epsilon_1(k))(\epsilon_1(p) \cdot \epsilon_1(k))\right] \right. \nonumber \\
& \left. + (k \cdot \epsilon_1(p))(k \cdot \epsilon_2(p))\left[(p \cdot \epsilon_1(k))^2-(p \cdot \epsilon_2(k))^2\right]\right\}\, , \\
g_7=&(G_1+G_2+2G_4)\left\{(k \cdot p)^2\left[(\epsilon_1(p) \cdot \epsilon_1(k))^2 - (\epsilon_2(p) \cdot \epsilon_1(k))^2+ (\epsilon_2(p) \cdot \epsilon_2(k))^2 - (\epsilon_1(p) \cdot \epsilon_2(k))^2\right]\right.\nonumber\\
&\left.-2(k \cdot p)(k \cdot \epsilon_2(p))\left[(p \cdot \epsilon_2(k))(\epsilon_2(p) \cdot \epsilon_2(k))- (p \cdot \epsilon_1(k))(\epsilon_2(p) \cdot \epsilon_1(k))\right]\right.\nonumber\\
&\left.+ 2(k \cdot p)(k \cdot \epsilon_1(p))\left[(p \cdot \epsilon_2(k))(\epsilon_1(p) \cdot \epsilon_2(k))- (p \cdot \epsilon_1(k))(\epsilon_1(p) \cdot \epsilon_1(k))\right]\right.\nonumber\\
& \left. + \left[(p \cdot \epsilon_1(k))^2- (p \cdot \epsilon_2(k))^2\right]\left[(k \cdot \epsilon_1(p))^2 - (k \cdot \epsilon_2(p))^2\right]\right\}\, ,\\
s_1=&\frac{(G_1+G_3+2G_4)}{(G_1+G_2+2G_4)}~g_1 \nonumber\\
=&(G_1+G_3+2G_4)\left\{(k \cdot p)^2\left[(\epsilon_2(p) \cdot \epsilon_1(k))(\epsilon_2(p) \cdot \epsilon_2(k))+(\epsilon_1(p) \cdot \epsilon_1(k))(\epsilon_1(p) \cdot \epsilon_2(k))\right]\right.\nonumber\\
&\left.-(k \cdot p)(k \cdot \epsilon_2(p))\left[(p \cdot \epsilon_2(k))(\epsilon_2(p) \cdot \epsilon_1(k))+(p \cdot \epsilon_1(k))(\epsilon_2(p) \cdot \epsilon_2(k))\right]\right.\nonumber\\
&\left.- (k \cdot p)(k \cdot \epsilon_1(p))\left[(p \cdot \epsilon_2(k))(\epsilon_1(p) \cdot \epsilon_1(k))+(p \cdot \epsilon_1(k))(\epsilon_1(p) \cdot \epsilon_2(k))\right]\right.\nonumber\\
& \left. + (p \cdot \epsilon_1(k))(p \cdot \epsilon_2(k))\left[(k \cdot \epsilon_1(p))^2+(k \cdot \epsilon_2(p))^2\right]\right\}\, ,\\
s_2=&\frac{(G_1+G_3+2G_4)}{(G_1+G_2+2G_4)}~g_3 \nonumber\\
=&(G_1+G_3+2G_4) \left\{(k \cdot p)^2\left[ (\epsilon_1(p) \cdot \epsilon_1(k))^2-(\epsilon_1(p) \cdot \epsilon_2(k))^2 +(\epsilon_2(p) \cdot \epsilon_1(k))^2-(\epsilon_2(p) \cdot \epsilon_2(k))^2\right] \right. \nonumber\\
& \left. +2(k \cdot p)(k \cdot \epsilon_1(p))\left[(p \cdot \epsilon_2(k))(\epsilon_1(p) \cdot \epsilon_2(k))-(p \cdot \epsilon_1(k))(\epsilon_1(p) \cdot \epsilon_1(k))\right] \right. \nonumber\\
&\left.+2 (k \cdot p)(k \cdot \epsilon_2(p)) \left[(p \cdot \epsilon_2(k))(\epsilon_2(p) \cdot \epsilon_2(k))-(p \cdot \epsilon_1(k))(\epsilon_2(p) \cdot \epsilon_1(k))\right]\right. \nonumber\\
& \left. + \left((k \cdot \epsilon_1(p))^2+(k \cdot \epsilon_2(p))^2\right)\left[(p \cdot \epsilon_1(k))^2-(p \cdot \epsilon_2(k))^2\right]\right\}\, ,
\end{align}
where $\epsilon_1(p)$ and $\epsilon_2 (p)$ denote the two independent transverse polarizations of a massless spin-1 particle.

\section{coefficients of Stokes parameters in the case of Euler-Heisenberg interaction} \label{appendix_B}
Here, we provide the explicit expression of the coefficients in Eqs. \eqref{eq:SSSPQ}, \eqref{eq:SSSPU}, and \eqref{eq:SSSPV}:
\begin{align}
f_1=&3(k \cdot p)^2\left[(\epsilon_2(p) \cdot \epsilon_1(k)) (\epsilon_2(p) \cdot \epsilon_2(k))+(\epsilon_1(p) \cdot \epsilon_1(k)) (\epsilon_1(p) \cdot \epsilon_2(k))\right]\nonumber\\
& -3(k \cdot p)(k \cdot \epsilon_2(p))\left[(p \cdot \epsilon_2(k))(\epsilon_2(p) \cdot \epsilon_1(k))+(p \cdot \epsilon_1(k))(\epsilon_2(p) \cdot \epsilon_2(k))\right] \nonumber\\
& -3(k \cdot p)(k \cdot \epsilon_1(p))\left[(p \cdot \epsilon_2(k))(\epsilon_1(p) \cdot \epsilon_1(k))+(p \cdot \epsilon_1(k))(\epsilon_1(p) \cdot \epsilon_2(k))\right]\nonumber\\
& + 3 (p \cdot \epsilon_1(k)) (p \cdot \epsilon_2(k))  \left[(k \cdot \epsilon_1(p))^2+(k \cdot \epsilon_2(p))^2\right]\, , \\
f_2=&3(k \cdot p)^2\left[(\epsilon_2(p) \cdot \epsilon_1(k)) (\epsilon_1(p) \cdot \epsilon_2(k))+(\epsilon_1(p) \cdot \epsilon_1(k)) (\epsilon_2(p) \cdot \epsilon_2(k))\right]\nonumber\\
&-3(k \cdot p)(k \cdot \epsilon_2(p))\left[(p \cdot \epsilon_2(k))(\epsilon_1(p) \cdot \epsilon_1(k))+(p \cdot \epsilon_1(k))(\epsilon_1(p) \cdot \epsilon_2(k))\right] \nonumber\\
&-3(k \cdot p)(k \cdot \epsilon_1(p))\left[(p \cdot \epsilon_2(k))(\epsilon_2(p) \cdot \epsilon_1(k))+(p \cdot \epsilon_1(k))(\epsilon_2(p) \cdot \epsilon_2(k)) \right. \nonumber\\
&\left.  +6(k \cdot \epsilon_1(p))(k \cdot \epsilon_2(p))(p \cdot \epsilon_1(k))(p \cdot \epsilon_2(k))\right]\, ,\\
f_3=&3(k \cdot p)^2\left[(\epsilon_1(p) \cdot \epsilon_1(k)) (\epsilon_1(p) \cdot \epsilon_2(k)) - (\epsilon_2(p) \cdot \epsilon_1(k)) (\epsilon_2(p) \cdot \epsilon_2(k))\right]\nonumber\\
&-3(k \cdot p)(k \cdot \epsilon_1(p))\left[(p \cdot \epsilon_1(k))(\epsilon_1(p) \cdot \epsilon_2(k)) + (p \cdot \epsilon_2(k))(\epsilon_1(p) \cdot \epsilon_1(k))\right] \nonumber\\
&+3(k \cdot p)(k \cdot \epsilon_2(p))\left[(p \cdot \epsilon_1(k))(\epsilon_2(p) \cdot \epsilon_2(k)) + (p \cdot \epsilon_2(k))(\epsilon_2(p) \cdot \epsilon_1(k)) \right] \nonumber\\
&-3 (p \cdot \epsilon_1(k)) (p \cdot \epsilon_2(k))[ (k \cdot \epsilon_2(p))^2 - (k \cdot \epsilon_1(p))^2] \, , \\
f_4=&3(k \cdot p)^2\left[(\epsilon_1(p) \cdot \epsilon_1(k))^2- (\epsilon_1(p) \cdot \epsilon_2(k))^2+(\epsilon_2(p) \cdot \epsilon_1(k))^2- (\epsilon_2(p) \cdot \epsilon_2(k))^2\right]\nonumber\\
&+6(k \cdot p)(k \cdot \epsilon_1(p))\left[(p \cdot \epsilon_2(k))(\epsilon_1(p) \cdot \epsilon_2(k))-(p \cdot \epsilon_1(k))(\epsilon_1(p) \cdot \epsilon_1(k))\right] \nonumber\\
&+6(k \cdot p)(k \cdot \epsilon_2(p))\left[(p \cdot \epsilon_2(k))(\epsilon_2(p) \cdot \epsilon_2(k))-(p \cdot \epsilon_1(k))(\epsilon_2(p) \cdot \epsilon_1(k)) \right] \nonumber\\
&+3\left((k \cdot \epsilon_1(p))^2+(k \cdot\epsilon_2(p))^2\right)\left((p \cdot \epsilon_1(k))^2-(p \cdot \epsilon_2(k))^2\right)\, , \\
f_5=& 6(k \cdot p)^2\left[(\epsilon_2(p) \cdot \epsilon_1(k)) (\epsilon_1(p) \cdot \epsilon_1(k))-(\epsilon_1(p) \cdot \epsilon_2(k)) (\epsilon_2(p) \cdot \epsilon_2(k))\right]\nonumber\\
&+6(k \cdot p)(k \cdot \epsilon_1(p))\left[(p \cdot \epsilon_2(k))(\epsilon_2(p) \cdot \epsilon_2(k))-(p \cdot \epsilon_1(k))(\epsilon_2(p) \cdot \epsilon_1(k))\right] \nonumber\\
&+6(k \cdot p)(k \cdot\epsilon_2(p))\left[(p \cdot \epsilon_2(k))(\epsilon_1(p) \cdot \epsilon_2(k))-(p \cdot \epsilon_1(k))(\epsilon_1(p) \cdot \epsilon_1(k)) \right] \nonumber\\
&+6(k \cdot \epsilon_1(p))(k \cdot \epsilon_2(p))\left((p \cdot \epsilon_1(k))^2-(p \cdot \epsilon_2(k))^2\right)\, , \\
f_6=&3(k \cdot p)^2\left[(\epsilon_1(p) \cdot \epsilon_1(k))^2- (\epsilon_2(p) \cdot\epsilon_1(k))^2-(\epsilon_1(p) \cdot \epsilon_2(k))^2 + (\epsilon_2(p) \cdot \epsilon_2(k))^2\right]\nonumber\\
&+6(k \cdot p)(k \cdot \epsilon_1(p))\left[(p \cdot \epsilon_2(k))(\epsilon_1(p) \cdot \epsilon_2(k))-(p \cdot \epsilon_1(k))(\epsilon_1(p) \cdot \epsilon_1(k))\right] \nonumber\\
&+6(k \cdot p)(k \cdot \epsilon_2(p))\left[(p \cdot \epsilon_1(k))(\epsilon_2(p) \cdot \epsilon_1(k))-(p \cdot \epsilon_2(k))(\epsilon_2(p) \cdot \epsilon_2(k)) \right] \nonumber\\
&+3 [(k \cdot \epsilon_1(p))^2-(k \cdot \epsilon_2(p))^2] \left((p \cdot \epsilon_1(k))^2-(p \cdot \epsilon_2(k))^2\right)\, .
\end{align}
\section{Stokes parameters for massive spin-1 particles} \label{appendix_C}
In this appendix, we briefly review the definition of Stokes parameters for massive spin-1 particles. As was shown in the text, one can parametrize the intensity and the polarization of massless photons using a $2\times 2$ polarization matrix involving four Stokes parameters: 
\beq
\rho_{ij}=\frac{tr(\rho)}{2}[ \mathbb{1}_2   + \bm{\sigma}\cdot \mathbf{P}]\, ,
\eeq
where $\mathbf{P}=(U, V, Q)/tr(\rho)$, $tr(\rho) = I$, and $\sigma_i$ are the Pauli matrices associated with the generators of the SU(2) group. However, when we want to describe \textit{massive} photons, it is more convenient to use an alternative representation of the polarization matrix that is made with the generators $\lambda_i$, $i=1,..., 8$ of the SU(3) group. In fact, the polarization matrix can be also written as \cite{Lakin-1955, Gomes-1981, Ramachandran-1980}
\beq \label{eq:reprS3}
\rho_{ij}=\frac{tr(\rho)}{3}\left[ \mathbb{1}_3 + \sum_{i=1}^{8} \lambda_i T_i \right]\, ,
\eeq
where $\rho_{ij}$ is a $3\times 3$ matrix, the generators $\lambda_i$ satisfy $tr(\lambda_i\lambda_j)=3\delta_{ij}$, and $T_i$ are eight parameters defined as $T_i=tr(\lambda_i\rho)/tr(\rho)$, which describe all the possible polarization states of massive photons. In fact, as discussed, e.g., in Ref. \cite{Ramachandran-1980}, the representation of Eq. \eqref{eq:reprS3} is sufficiently general to describe not only ``physical photons'' which are only transverse, but also massive photons, admitting longitudinal polarization states. As an example, consider a spin-1 particle moving along the $z$ axis. In this case, taking $ \vec \epsilon = (\epsilon_x, \epsilon_y, \epsilon_z)$ as its polarization vector, it is possible to show that the corresponding polarization matrix can be written as \cite{Ramachandran-1980}
\beq
 \rho_{ij}=\frac{1}{2}\left(
                  \begin{array}{ccc}
                  \epsilon_x \epsilon_x^* + \epsilon_y \epsilon_y^*  + 2 \textrm{Im}(\epsilon_x \epsilon_y^*) &- \sqrt 2 (\epsilon_x + i \epsilon_y) \epsilon_z^*  & - \epsilon_x \epsilon_x^* + \epsilon_y\epsilon_y^*  - 2 i \textrm{Re}(\epsilon_x \epsilon_y^*) \\
                     - \sqrt 2 (\epsilon_x^* + i \epsilon_y^*) \epsilon_z & 2 \epsilon_z \epsilon_z^*  & \sqrt 2 (\epsilon_x^* + i \epsilon_y^*) \epsilon_z  \\
                    -\epsilon_x \epsilon_x^* + \epsilon_y \epsilon_y^*  + 2 i \textrm{Re}(\epsilon_x \epsilon_y^*)) & \sqrt 2 (\epsilon_x - i \epsilon_y) \epsilon_z^*  &   \epsilon_x \epsilon_x^* + \epsilon_y \epsilon_y^*  - 2 \textrm{Im}(\epsilon_x \epsilon_y^*) \\
                   \end{array}
                 \right)\, .\label{eq:reprS3e}
\eeq
Now, by matching Eq. \eqref{eq:reprS3} with Eq. \eqref{eq:reprS3e}, we get the following definition of the parameters $T_i$ in terms of the photon polarization vector:
\begin{align}
T_1 =& - \frac{1}{\sqrt 2} \textrm{Re}((\epsilon_x + i \epsilon_y) \epsilon_z^*)\, ,\\
T_2 =& - \frac{1}{\sqrt 2} \textrm{Im}((\epsilon_x + i \epsilon_y) \epsilon_z^*) \, ,\\
T_3 =&  \frac{1}{2} \sqrt{\frac{3}{2}} (1 - 3 \epsilon_z \epsilon_z^* + 2 \textrm{Im}(\epsilon_x \epsilon_y^*)) \, ,\\
T_4 =& -  \frac{1}{2} (\epsilon_x \epsilon_x^* - \epsilon_y \epsilon_x^*) \, , \\
T_5 =& - \textrm{Re}(\epsilon_x \epsilon_y^*) \, , \\
T_6 =& \frac{1}{\sqrt 2}  \textrm{Re}((\epsilon_x^* + i \epsilon_y^*) \epsilon_z) \, , \\
T_7 =& \frac{1}{\sqrt 2}  \textrm{Im}((\epsilon_x^* + i \epsilon_y^*) \epsilon_z) \, , \\
T_8 =& \frac{3}{2 \sqrt 2} \left[ \epsilon_z \epsilon_z^* + 2 \textrm{Im}(\epsilon_x \epsilon_y^*) - \frac{1}{3}  \right] \, .
\end{align}
For massless photons with no longitudinal polarization states, we have necessarily $\epsilon_z = 0$. As a consequence, we get $T_1 = T_2 = T_6 = T_7 = 0$, and $T_3$ is related to $T_8$. Thus, only three independent degrees of freedom remain, which correspond to the usual $Q$, $U$, and $V$ Stokes parameters. However, if we include in the picture the possibility to have longitudinal polarization states, we need to describe the polarization state of a spin-1 particle with the $T_i$ parameters that play the role of \textit{generalized} Stokes parameters.
\section{coefficients of Stokes parameters in the case of general photon-spin-1 interaction} \label{appendix_D}
 Here, we provide the explicit expression of the coefficients in Eqs. \eqref{eq:massiveQ}, \eqref{eq:massiveU}, and \eqref{eq:massiveV}:
 \begin{align}
L_0=& (4A_4 (k\cdot p) - k p (A_1+2A_2)+4 m_D^2 A_3) \left\{(k \cdot p)^2\left[(\epsilon_2(p) \cdot \epsilon_1(k)) (\epsilon_2(p) \cdot \epsilon_2(k))+(\epsilon_1(p) \cdot \epsilon_1(k)) (\epsilon_1(p) \cdot \epsilon_2(k)) \right.\right.\nonumber\\
&\left.\left. + (\epsilon_\ell(p) \cdot \epsilon_1(k)) (\epsilon_\ell(p) \cdot \epsilon_2(k))\right] -(k \cdot p)(k \cdot \epsilon_\ell(p))\left[(p \cdot \epsilon_2(k))(\epsilon_\ell(p) \cdot \epsilon_1(k))+(p \cdot \epsilon_1(k))(\epsilon_\ell(p) \cdot \epsilon_2(k))\right] \right. \nonumber\\
& \left. -(k \cdot p)(k \cdot \epsilon_2(p))\left[(p \cdot \epsilon_2(k))(\epsilon_2(p) \cdot \epsilon_1(k))+(p \cdot \epsilon_1(k))(\epsilon_2(p) \cdot \epsilon_2(k))\right] \right. \nonumber\\
& \left. -(k \cdot p)(k \cdot \epsilon_1(p))\left[(p \cdot \epsilon_2(k))(\epsilon_1(p) \cdot \epsilon_1(k))+(p \cdot \epsilon_1(k))(\epsilon_1(p) \cdot \epsilon_2(k))\right]\right.\nonumber\\
& \left.+  (p \cdot \epsilon_1(k)) (p \cdot \epsilon_2(k))  \left[(k \cdot \epsilon_\ell(p))^2+(k \cdot \epsilon_1(p))^2+(k \cdot \epsilon_2(p))^2\right] \right\}\, , \\
M_0 =& (4A_4 (k\cdot p) - k p (A_1+2A_2)+4 m_D^2 A_3) \left\{(k \cdot p)^2\left[(\epsilon_2(p) \cdot \epsilon_1(k))^2 - (\epsilon_2(p) \cdot \epsilon_2(k))^2+(\epsilon_1(p) \cdot \epsilon_1(k))^2 - (\epsilon_1(p) \cdot \epsilon_2(k))^2 \right.\right.\nonumber\\
& \left.\left. + (\epsilon_\ell(p) \cdot \epsilon_1(k))^2 - (\epsilon_\ell(p) \cdot \epsilon_2(k))^2\right] +2(k \cdot p)(k \cdot \epsilon_\ell(p))\left[(p \cdot \epsilon_2(k))(\epsilon_\ell(p) \cdot \epsilon_2(k))-(p \cdot \epsilon_1(k))(\epsilon_\ell(p) \cdot \epsilon_1(k))\right] \right. \nonumber\\
& \left. +2(k \cdot p)(k \cdot \epsilon_2(p))\left[(p \cdot \epsilon_2(k))(\epsilon_2(p) \cdot \epsilon_2(k))-(p \cdot \epsilon_1(k))(\epsilon_2(p) \cdot \epsilon_1(k))\right] \right. \nonumber\\
& \left. +2(k \cdot p)(k \cdot \epsilon_1(p))\left[(p \cdot \epsilon_2(k))(\epsilon_1(p) \cdot \epsilon_2(k)) -(p \cdot \epsilon_1(k))(\epsilon_1(p) \cdot \epsilon_1(k))\right]\right.\nonumber\\
& \left.+  \left[(p \cdot \epsilon_1(k))^2- (p \cdot \epsilon_2(k))^2 \right]  \left[(k \cdot \epsilon_\ell(p))^2+(k \cdot \epsilon_1(p))^2+(k \cdot \epsilon_2(p))^2\right] \right\}\, ,
\end{align}
where $m_D$ denotes the mass of the spin-1 particle and $\epsilon_\ell (p)$ its longitudinal polarization.
\end{appendices}

\begingroup 
\makeatletter
\let\ps@plain\ps@empty
\makeatother
\bibliography{reference}
\endgroup

\end{document}